\documentclass[10pt]{article}

\usepackage{graphicx}
\usepackage{cite}
\usepackage{multirow}
\usepackage{dsfont}
\usepackage{verbatim}
\usepackage{amsmath}

\usepackage[labelfont=bf,labelsep=period,justification=raggedright]{caption}

\def\`#1{\textbf{#1}}
\def\*#1{\mathbf{#1}}
\def\!#1{\mathcal{#1}}
\def\^#1{\left\langle{#1}\right\rangle}
\def\TPB{\mathcal{TPB}}
\def\TPBN{\mathcal{TPBN}}
\def\N{\mathcal{N}}
\def\Ga{\mathcal{G}a}
\def\z_k{\langle z_k \rangle}
\def\({\left(}
\def\){\right)}

\DeclareMathOperator*{\argmax}{arg\,max}
\begin{document}

\begin{flushleft}
{\Large
\textbf{A latent factor model with a mixture of sparse and dense factors to  model gene expression data with confounding effects}
}
% Insert Author names, affiliations and corresponding author email.
\\
Chuan Gao$^{1}$, 
Christopher D Brown$^{2}$, 
Barbara E Engelhardt$^{1,3,\ast}$
\\
\bf{1} Institute for Genome Sciences \& Policy, Duke University, Durham, NC, USA
\\
\bf{2} Department of Genetics, University of Pennsylvania, Philadelphia, PA, USA
\\
\bf{3} Department of Biostatistics \& Bioinformatics and Department of Statistical Science, Duke University, Durham, NC, USA
\\
$\ast$ E-mail: barbara.engelhardt@duke.edu
\end{flushleft}

%\end{document}

\begin{abstract}
%%\vspace{-0.2cm}
%Sparse latent factor models have been extremely useful in extracting low dimensional, interpretable features and feature relationships from high dimensional data. Interpretability is especially important for problems in genome sciences, where high-dimensional data are generated at an exploding rate, but analytic tools for these data have not kept pace. 
One important problem in genome science is to determine sets of co-regulated genes based on measurements of gene expression levels across samples, where the quantification of expression levels includes substantial technical and biological noise. To address this problem, we developed a Bayesian sparse latent factor model that uses a three parameter beta prior to flexibly model shrinkage in the loading matrix. By applying three layers of shrinkage to the loading matrix (global, factor-specific, and element-wise), this model has non-parametric properties in that it estimates the appropriate number of factors from the data. We added a two-component mixture to model each factor loading as being generated from either a sparse or a dense mixture component; this allows dense factors that capture confounding noise, and sparse factors that capture local gene interactions. We developed two statistics to quantify the stability of the recovered matrices for both sparse and dense matrices. We tested our model on simulated data and found that we successfully recovered the true latent structure as compared to related models. We applied our model to a large gene expression study and found that we recovered known covariates and small groups of co-regulated genes. We validated these gene subsets by testing for associations between genotype data and these latent factors, and we found a substantial number of biologically important genetic regulators for the recovered gene subsets.
\end{abstract}

%\begin{keyword}[class=AMS]
%\kwd[Primary ]{60K35}
%\kwd{60K35}
%\kwd[; secondary ]{60K35}
%\end{keyword}

%%\vspace{-0.2cm}
\section{Introduction}
%\vspace{-0.2cm}

% paragraph 1: problem of gene expression analysis, identifying trans-eQTLs
Fast evolving experimental techniques for assaying genomic data have enabled the generation of large scale gene expression and genotype data at an unprecedented pace~\cite{Wang:RNAseq2009,Hapmap:2008}. Studies to find genetic variants that regulate gene expression levels (called expression quantitative trait loci, or eQTLs), are now possible~\cite{Cookson:2009,gilad:2008}. However, due to the complicated nature of experimental assays to quantify cellular traits, substantial technical noise and biological covariates may confound measurements of gene expression levels. These confounding effects include batch effects~\cite{Leek:2007,Friguet:2009,Stegle:2010,Listgarten:2010,Yang:2013}, latent population structure among the samples~\cite{Price:2006,Pritchard:2000,Runcie:2013}, and biological covariates, including age, sex, or body mass index (BMI). %allelic heterogeneity~\cite{West:2007,Brem:2002,Brown:arXiv2012} and pleiotropy~\cite{Featherstone:2002,Stearns:2010}. 

\begin{figure}[!htb]
\begin{center}   
\includegraphics[width=1\linewidth]{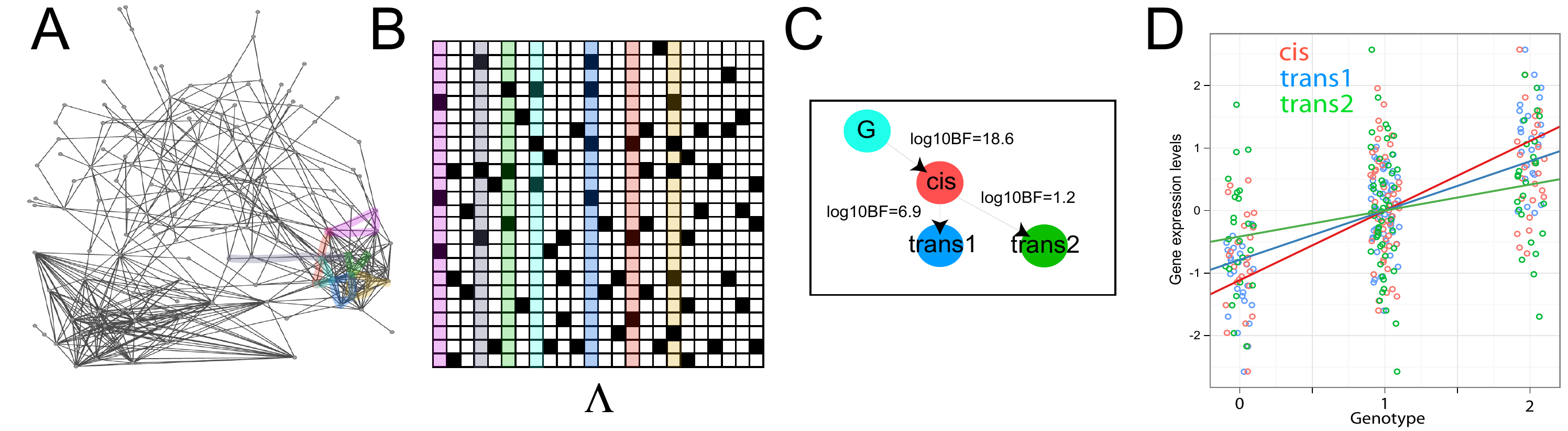}

\caption{{\bf Small gene network modules and their role in identifying pleiotropic eQTLs.} Panel A: A large gene network shown as multiple nodes (genes) connected by edges (estimated using partial correlation between gene expression levels), with small sub-networks highlighted. Panel B: Each column of this matrix represents a gene cluster with black elements denoting included genes in this latent factor. The colored columns correspond to the colored subsets in Panel A. Panel C: A directed network, including a $cis$ regulatory genetic variant ($G$) that regulates a gene in cis (`cis') and two $trans$ regulatory genetic variants. Panel D: the x-axis is the three states of the single SNP, and the y-axis is the gene expression level for that transcription factor across samples. The slope of the line connecting the means for each simulated gene is the effect size of the SNP on the transcription levels of that gene.
} \label{fig:gene-network}
\end{center}
\end{figure}

The most replicable and numerous eQTL associations that have been identified in humans are those for which the single nucleotide polymorphism (SNP) is within the \emph{cis} region of, or local to, the associated gene~\cite{Stranger:2005,Mangravite:2013}. In practice, eQTL analyses are conducted by testing each genetic variant for an additive association with only the genes in \emph{cis}, or local, which helps to alleviate some of the burden imposed by multiple testing~\cite{Freyhult:2010,Jiang:2004}. The biological reality is that genes cannot manifest their function alone; instead, genes tend to work together to achieve biological functions (Figure~\ref{fig:gene-network}A)~\cite{Gasch:2002,Allocco:2004,Hobert:2008}. Furthermore, a SNP that regulates a gene in \emph{cis} that, in turn, drives the expression levels of other genes, such as a transcription factor, may appear to co-regulate a subnetwork of genes (possibly in \emph{trans}; Figure~\ref{fig:gene-network}C,D). Methods that identify small, co-regulated groups of genes provide important information to a downstream eQTL analysis, enabling genetic variants that regulate multiple genes (\emph{pleiotropic} eQTLs) to be identified (Figure~\ref{fig:gene-network}B,D). 

% paragraph 2: problem specification
The most effective method to control for confounding effects in gene expression assays in order to have power to identify eQTLs remains an open question. Confounding effects are often controlled by estimating principal components (PCs) of the gene expression matrix and removing the effects of the initial PCs before downstream analysis on the normalized residuals~\cite{Pickrell:2010,Mangravite:2013}; 
the downside of this two-step procedure is that it is possible that some of the sparse signal is removed in the first step~\cite{Stegle:2010,Goldinger:2013}. We address this problem by developing a Bayesian latent factor model to identify a large number of sparse gene clusters, where individual signals are perturbed by unobserved confounding noise. In this latent factor model, small clusters of co-regulated genes are captured by a large number of sparse factors. To jointly model and implicitly control for confounding noise, our model includes a two-component mixture that allows each factor loading to be regularized by either a sparsity-inducing prior or an alternative prior that does not induce sparsity, where the probability of a factor loading being sparse or dense is estimated from the data. 

% paragraph 3: current paradigm
Latent factor models, and sparse latent factor models in particular, are a common and effective statistical methodology for identifying interpretable, low dimensional structure within a high dimensional matrix, and have frequently been used to identify latent structure in gene expression data~\cite{pournara:2007,Blum:2010,Lucas:2010,Parts:2011}. This approach assumes that the gene expression levels for each gene can be described by a linear combination of latent factors, and that the random noise in this matrix is approximately normal; thus each sample is modeled as being drawn from a multivariate normal distribution with a diagonal covariance matrix across genes, where the mean parameter is a linear combination of latent factors with a normal prior, and the variance term is estimated for each feature separately. Latent factor models assume that the total variation within the matrix can be partitioned into covariation among genes and variation specific to genes. This implies that a set of genes with correlated gene expression levels will contribute substantially to (have a substantial loading on) a single factor, because this co-variability will contribute to the overall variability in the matrix. In the setting of gene expression data, sparsity has often been imposed on the loading matrix to facilitate this clustering interpretation: genes with zero contribution to a factor are not included in the associated gene cluster~\cite{Carvalho:BFRM}. 

% paragraph 4: What we do
In this work, we develop a flexible Bayesian sparse latent factor model, and we extend this sparse factor model to capture both sparse and dense latent factors by including a two-component mixture of priors on the loading matrix. We use the flexible three parameter beta ($\mathcal{TPB}$) prior to induce local (element-specific), factor-specific, and global shrinkage within the loading matrix~\cite{Armagan:GeneralBeta}. We then add a two-component mixture  on the parameters of the factor-level three parameter beta prior to jointly model sparse and dense latent structure. While this model draws upon ideas in our previous work in sparse factor analysis~\cite{Engelhardt:2010,Gao:SFA}, the main contributions of this work are that i) we adapt the Bayesian two group regularization framework for regression~\cite{Polson:Global-Local,Efron:2008} to latent factor models in a natural way to create a flexible sparse latent factor model with desirable non-parametric and computational properties, and ii) we take advantage of this flexibility by jointly modeling sparse and dense factors. We believe that this sparse latent factor model will have broad utility in Bayesian statistics.% might add a few citations to other papers in vision, econ, etc. that use sfa type methods. 

A general difficulty when working with latent factor models is that, in the basic model, the factors and loadings are only identifiable up to orthogonal rotation, scaling, and label switching~\cite{Kaiser:varimax}. We would like to develop metrics with which to compare both sparse and dense matrices in order to evaluate convergence in parameter estimates and to quantify the similarity of the recovered matrices and the underlying structure. These metrics must be robust to these invariances to be useful in this setting. While sparsity in the loading matrix facilitates rotational identifiability and enables more direct comparisons across fitted sparse latent factors, dense factors are not as trivially comparable because of this rotation invariance. In order to address these issues of comparison, we developed two statistics to quantify the stability across estimated factors and factor loading vectors that are sparse (contain zeros) and dense (do not contain zeros). Both statistics are invariant to label switching and scale. In addition, the dense matrix stability statistic is rotation invariant.

% paragraph 5: layout of paper
This paper is organized as follows. Section~\ref{sec:bkg-SFA} provides a general background to sparse factor analysis to motivate our formulation of a Bayesian sparse latent factor model. Section~\ref{sec:SFA} specifies our  factor model with the $\mathcal{TPB}$ prior and the equivalent model in terms of a simple hierarchical gamma prior. Section~\ref{sec:mix-SFA} extends the model to include a mixture of sparse and dense factors. The parameters are estimated using an approximate EM algorithm outlined in Section~\ref{sec:EM-main} and Appendix~\ref{sec:EM}. We motivate and describe our stability statistics in Section~\ref{sec:robust}. To evaluate the performance of our model, we simulated data and compared our model to related methods based on these simulations (Section~\ref{sec:sim-result}). In Section~\ref{sec:real}, we applied this model to real gene expression data on $480$ samples and $8,718$ genes, revealing interesting patterns in gene expression and confounding factors. Using these factors, we identify relevant genetic associations for the subsets of co-regulated genes.   

\section{Bayesian sparsity and latent factor models}\label{sec:bkg-SFA}

Factor analysis has been used in a variety of settings to extract useful low dimensional features from high dimensional data~\cite{Steward:1981,pournara:2007,Blum:2010,Lucas:2010,Parts:2011}. We begin with a basic factor analysis model~\cite{rubin:1982,West:2003}, $\*Y = \*X \*\Lambda+ \*\epsilon$, with $\*Y \in \Re^{n \times p}$, $\*X \in \Re^{n \times K}$, $\*\Lambda \in \Re^{K \times p}$, and $\*\epsilon \in \Re^{n\times p}$, $\epsilon_{ij} \sim {\mathcal N}(0,\psi_j)$, where $n$ and $p$ correspond respectively to the number of samples and the number of genes and, in practice, $n \ll p$.  
To ensure conjugacy, the loading matrix $\*\Lambda$ and the latent factors $\*X$ have normal priors. %Under this model, the covariance among the features is approximated by $\*\Lambda^T \*\Lambda$.
This basic factor analysis model has a number of drawbacks: the latent factors and corresponding loadings are unidentifiable with respect to orthogonal rotation and scaling, and it is difficult to select the dimension of the latent factors, which is fixed \emph{a priori}. One solution to addresses rotational invariance is to induce sparsity in the loading matrix, which allows for identifiability in the estimated matrices when the latent space is sufficiently sparse~\cite{Engelhardt:2010}. %With sufficient regularization, model selection criteria appear to be more effective at selecting the dimensionality of the latent space because the model does not overfit to the same extent as a non-sparse model~\cite{Bhat:biometrika}. % What does this last sentence add to this paragraph? **answer: fixed

There are currently a number of ways to regularize the latent parameter space. Sparse principle component analyses (PCA) have been described~\cite{Zou:2004,Witten:2009}, related to latent factor models through a probabilistic PCA framework~\cite{Roweis:1998,Tipping:1999,Engelhardt:2010}; for example, sparse principle components analysis (SPCA) uses an $\ell_1$ penalty to induce sparsity on the PCs~\cite{Zou:2004,Witten:2009}. We choose to work in the Bayesian context with latent factor models, and consider a sparsity-inducing prior on the factor loading matrix $\Lambda$. This sparsity-inducing prior should have substantial mass around zero to provide strong shrinkage near zero, and also have heavy tails to allow signals to escape strong shrinkage~\cite{Polson:Global-Local,Carvalho:HorseShoe}. In the context of sparse regression, there have been a number of proposed solutions, including a student's t-distribution, the horseshoe prior, the normal-gamma prior, and the Laplace prior~\cite{Tipping:2001,Carvalho:HorseShoe,Brown:normal-gamma,Trevor:Bay-lasso}.
Sparse factor analysis models have taken advantage of some of these sparsity-inducing priors ~\cite{Pruteanu:2011,Bhat:biometrika,Engelhardt:2010}.
In particular, a number of sparse factor models for use in biological applications have included some form of the Student's t-distribution, also known as automatic relevance determination (ARD)~\cite{Tipping:2001,Neal:1996}, as a prior on the variance terms of the elements of the factor loading matrix~\cite{Fokoue:04,Engelhardt:2010,Pruteanu:2011}. 
The sparse Bayesian infinite factor model (SBIF)~\cite{Bhat:biometrika}
introduces increasingly stronger shrinkage across the loading vectors
using a multiplicative gamma prior. The Infinite Sparse Factor Analysis model (ISFA)~\cite{Knowles:2007,Knowles:2011} extends the Indian Buffet Process to select the number of latent factors in the sparse loading matrix. In these two models, as the proportion of variance explained by the factors decreases, the proportion of zeros in the factor loadings, in theory, increases, enabling a finite number of factors to be recovered from a model with an infinite number of underlying factors. Two flaws in this construction are that i) sparsity and PVE may not be well correlated in the latent space we are modeling and ii) PVE may not be a monotone decreasing function.

In this sparse factor analysis context, most approaches to inducing sparsity have applied shrinkage through a single parameter (generally, the variance of the factor loading matrix elements) on all loading parameters, which may sacrifice small signals to achieve high levels of sparsity.  This behavior has been labeled the \emph{one group} solution to inducing sparsity, because it effectively considers both signal and noise in a single group and regularizes them the same way~\cite{Efron:2008}. In contrast, the \emph{two groups} solution models noise and signal differently, strongly shrinking noise to zero but allowing signals to escape extreme shrinkage~\cite{Polson:Global-Local}.

The canonical \emph{two groups} solution in the Bayesian context is the so-called `spike-and-slab' prior, which induces sparsity using a two-component mixture model including a point mass at zero and a normal distribution~\cite{Mitchell:1988,George:1993}. % the horseshoe paper is not relevant here!
The components that are noise are effectively removed from the model through the point mass at zero, while the signals are regularized using the normal distribution but remain in the model; this approach additionally allows an explicit posterior distribution on the inclusion probability of each component~\cite{Carvalho:BFRM}.
In the factor model framework, a spike-and-slab prior can be put on each element of the loading matrix, as in the Bayesian factor regression model (BFRM)~\cite{Carvalho:BFRM}. 
Unfortunately, there is no closed form solution for the parameter estimates, because of the mixture component, and so MCMC is most generally used to estimate the parameters~\cite{Carvalho:BFRM}. Because the parameter space for $m$ components includes $2^m$ configurations, this is computational intractable for large matrices~\cite{Wagner:2011}.
These continuous sparsity-inducing priors all have the property that they impose strong shrinkage around zero but have sub-exponential tails, which allow signals to escape shrinkage. Because of these properties, these types of priors have been described as the `one-group answer to the original two-groups question'~\cite{Polson:Global-Local}. 

% note to self (bee): at some point, bring up dictonary learning.

In this work, we use a three parameter beta ($\mathcal{TPB}$) distribution~\cite{Armagan:GeneralBeta} to encourage sparsity in the elements of the factor loading matrix by shrinking their variance term. $\mathcal{TPB}(a,b,\phi)$ is a generalized form of the Beta distribution, with the third parameter $\phi$ further controlling the shape of the density. It has been shown that a linear transformation of the beta distribution, producing the inverse beta distribution or the \emph{horseshoe} prior, has desirable shrinkage properties in sparse regression~\cite{Carvalho:HorseShoe}. A linear transformation of the $\mathcal{TPB}$ distribution can be used to mimic the inverse beta distribution, with the inverse beta variable scaled by $\phi$. The $\mathcal{TPB}$ distribution can also replicate other distributions, including the Strawderman-Berger prior~\cite{Armagan:GeneralBeta}. The $\mathcal{TPB}$ produces a $\mathcal{TPB}$-normal ($\TPBN$) distribution when coupled with the normal distribution, where, for $a=1$ and $\phi=1$, this is equivalent to the normal-exponential-gamma distribution (NEG, Table~\ref{Tab:TPB-dist})~\cite{Armagan:GeneralBeta}.
The $\mathcal{TPB}$ is thus appealing as a prior because it can recapitulate the sparsity-inducing properties of continuous one group priors, including the horseshoe, but it is also flexible enough to recapitulate other types of priors including some that do not induce sparsity (Table~\ref{Tab:TPB-dist}). 
\setlength{\tabcolsep}{1em}
\begin{table}[h]
\caption{Effect of different parameter settings for $\TPB(a,b,\phi)$ on the shrinkage imposed by this prior.\label{Tab:TPB-dist}}
\begin{center}
\begin{tabular}{lccc}\hline
&\multicolumn{3}{c}{$\phi$}\\\cline{2-4}
&1&$<1$&$>1$\\\cline{2-4}
$a=b=\frac{1}{2}$& horseshoe  & \multirow{3}{*}{strong} & \multirow{3}{*}{weak}\\\cline{2-2}
$a=1,b=\frac{1}{2}$& Strawderman-Berger  &   & \\\cline{2-2}
$\TPBN$ for $a=1$& NEG  &  & \\\cline{2-4}
$a\uparrow$ and $b\downarrow$ & weak & variable & weak\\\cline{2-4}
$a\downarrow$ and $b\uparrow$ & strong & strong & variable \\\hline
\end{tabular}
\end{center}
\end{table} 

We build a sparse factor model using this sparsity-inducing prior following recent work in Bayesian regression~\cite{Polson:Global-Local}. In the regression context, a two groups model is achieved by setting the variance term for the regression coefficients to a scale mixture of normals:
\begin{eqnarray}
\beta_j | \lambda_j, \tau &\sim& \mathcal{N}(0, \tau^2\lambda^2_j)\\
\lambda_j &\sim& \pi(\lambda_j)\\
(\tau^2,\phi^2) &\sim& \pi(\tau, \phi),
\end{eqnarray}
where $\pi$, with a one group prior, is on the local variance component $\lambda_j$, and the same distribution is on the global variance component $\tau$. This simple model exhibits two groups behavior, given the proper distributions for $\tau$ and $\phi$, because $\tau$ effectively shrinks all of the regression coefficients to $0$, then $\lambda_j$, which is allowed to be very large through a heavy-tailed distribution, rescues individual signals~\cite{Polson:Global-Local} by scaling the global shrinkage parameter $\tau$ that is very small. 

To adapt this approach to the setting of latent factor models, we added an additional layer of shrinkage to each individual factor, which maintains the global-local-type model selection from the regression context, but allows factor-specific behavior. This creates, in effect, a three groups model, where signal and noise are modeled in a factor-specific way. In particular, a global parameter heavily shrinks all signals through the loading matrix toward zero, a factor-specific parameter rescues specific factors from global shrinkage, and a local parameter enables within-factor sparsity by shrinking individual elements of a factor. Each of the three layers serves a critical role: global regularization creates a non-parametric effect of removing factors from the model that are not necessary, factor-specific regularization identifies factors that will be included in the model, and local regularization enables sparsity, or model selection, within those selected factors. We impose regularization at all three levels of the loading matrix using the $\mathcal{TPB}$ prior because it is continuous and flexible.

Recent work has produced a strong result in the Bayesian sparse factor model setting that, using specific local-global shrinkage priors, one obtains the minimax optimal rate of posterior concentration up to a log factor; this work is the first asymptotic justification for global-local approaches to Bayesian sparse factor analysis~\cite{Pati2012}. Although we use a different heavy-tailed local prior, this work motivates our general approach to Bayesian sparse factor analysis. 

Our sparse latent factor model has a straightforward posterior distribution for which point estimates of the parameters are computed using expectation maximization (EM), making it computationally tractable via the careful application of this continuous distribution. Particularly in the \emph{dictionary learning} setting of identifying an \emph{overcomplete}, or $K > n$, number of factors that may individually contribute minimally to the variation in the response matrix, this approach to inducing sparse in latent factor models is statistically and computationally well motivated~\cite{Elad:2006KSVD}.

%\vspace{-0.2cm}
\section{Bayesian sparse factor model via $\mathcal{TPB}$}
\label{sec:SFA}

We define a Bayesian factor analysis model in the following way:
\begin{eqnarray}
\*Y &=&\*X \*\Lambda +  \*\epsilon \\\label{eq:model}
\*X_i &\sim& \mathcal{N}(0, \*I_K) \\
\*\epsilon_j &\sim& \mathcal{N}(0,\Psi_j),
\end{eqnarray}
where $\*Y \in \Re^{n \times p}$ is the matrix of observed variables, $\*X \in \Re^{n \times K}$ is the factor matrix with $K$ factors, $\*\Lambda \in \Re^{K \times p}$ is the loading matrix, and $\*\epsilon \in \Re^{n \times p}$ is the residual error matrix. We assume $\*\Psi = diag(\psi_1,\dots,\psi_p)$ is diagonal (but the diagonal elements are not necessarily the same). In this model, the covariance among the $p$ features in $\*Y$ is captured in $\*\Lambda^T\*\Lambda$. For the latent factors in $\*X$, we follow the usual convention by giving each element a standard normal prior, where $\*I_K$ is the $K\times K$ identity matrix.

\begin{figure}[!htb] 
\begin{center}   
\includegraphics[width=1\linewidth]{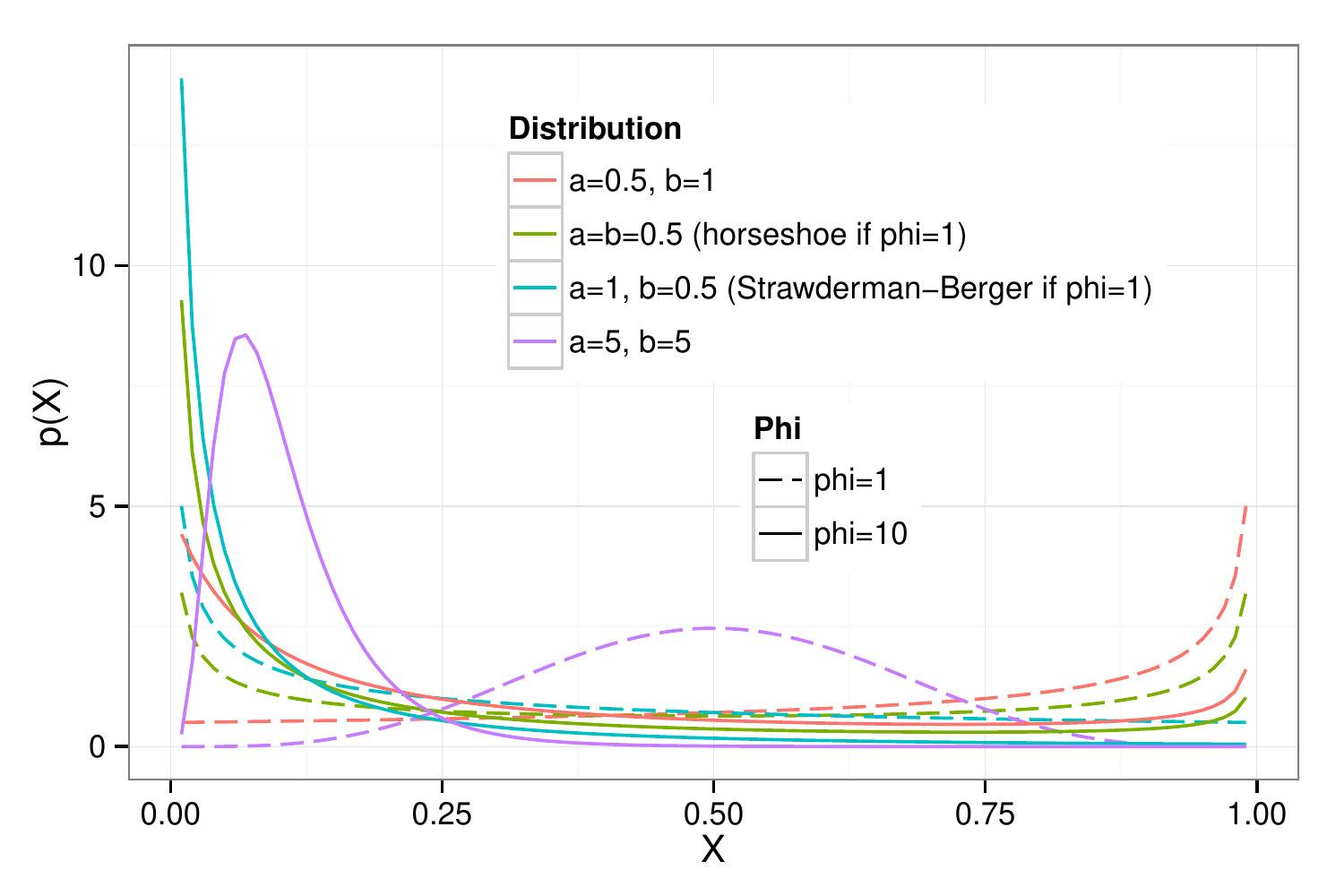}
\caption{{\bf The pdf of different parameterizations of the three parameter beta distribution.} The combination of a color and a line type identify alternative parameterizations and associated probability density functions (pdf; y-axis) on $x\in(0,1)$ (x-axis).}\label{fig:dist}
\end{center}
\end{figure}

To induce sparsity in the factor loading matrix $\*\Lambda$, we use the three parameter beta distribution parameterized to have a sparsity-inducing effect~\cite{Armagan:GeneralBeta}. The three parameter beta distribution has the following form:
\begin{equation}
f(x:a,b,\phi)=\frac{\Gamma(a+b)}{\Gamma(a)\Gamma(b)}\phi^bx^{b-1}(1-x)^{a-1}\{1+(\phi-1)x\}^{-(a+b)},
\end{equation}
for $x\in (0,1)$, $a>0$, $b>0$ and $\phi>0$. % can we visualize this somehow? Do we need a figure? I like that idea better than the table. 
% Is it (0,1) or [0,1]? can you check?
We put the $\TPB$ prior on the variance of each element $\Lambda_{k,j}$ of the loading matrix $\*\Lambda$, creating the following hierarchical structure:
\begin{eqnarray}
\varrho &\sim& \mathcal{TPB}(e,f,\nu).\label{spec:varrho}\\
\zeta_k &\sim& \mathcal{TPB}(c,d,\frac{1}{\varrho}-1)\label{spec:zeta}\\
\varphi_{k,j} &\sim& \mathcal{TPB}(a,b,\frac{1}{\zeta_k}-1)\label{spec:varphi}\\
\Lambda_{k,j} &\sim& \mathcal{N}(0,\frac{1}{\varphi_{k,j}}-1)\label{spec:lam}
\end{eqnarray}
This specification provides three layers of shrinkage on the sparse loading matrix: $\varphi_{k,j}$ provides local shrinkage for each element by shrinking the variance term of the normal prior; $\zeta_k$ controls the shrinkage specific to each factor $k$; $\varrho$ shrinks all elements of the matrix globally.
This model captures different shrinkage scenarios depending on $\mathcal{TPB}$ parameters at each level $a,b,c,d,e,f$ (Figure~\ref{fig:dist}). We estimate the third term of the factor-specific and local $\TPB$ priors, $\varrho,\zeta_k$, from the data. % let's talk about this sentence. I'm not sure it's right on a number of levels.

By tuning parameters $a,b,c,d,e,f$ and $\nu$, we apply more or less shrinkage on the sparse loading matrix $\*\Lambda$. In practice, we set $a=b=c=d=e=f=0.5$ to recapitulate the horseshoe prior at all three levels. % is this true independent of the third parameter?

\subsection{Equivalent model via gamma priors}\label{sec:str-SFA}
%\vspace{-0.2cm}
We transform the parameter $\varphi$ to $\theta=\frac{1}{\varphi}-1$, and we find that the following relationship holds~\cite{Armagan:GeneralBeta}:
\begin{equation}
\label{eq:gamma_beta_relation}
\varphi \sim \mathcal{TPB}(a,b,\nu) \Leftrightarrow \frac{\theta}{\nu} \sim Be^{\prime}(a,b) \Leftrightarrow \theta \sim \mathcal{G}a(a,\delta) \; \mathrm{ and } \; \delta \sim \mathcal{G}a(b,\nu),
\end{equation}
where $Be^{\prime}(a,b)$ and $\mathcal{G}a$ indicate an inverse beta and a gamma distribution, respectively. For concreteness, we define the inverse beta distribution as follows:
\begin{equation}
\label{eq:inv_beta}
f(x;\alpha,\beta)=\frac{x^{\alpha-1}(1+x)^{-\alpha-\beta}}{B(\alpha,\beta)},
\end{equation}
where $B(\cdot,\cdot)$ is the beta function. We apply this transformation to Equations~\ref{spec:varrho}, \ref{spec:zeta}, \ref{spec:varphi}, and \ref{spec:lam}, specifically, the variance terms $\theta_{k,j}=\frac{1}{\varphi_{k,j}}-1$ and $\phi_k=\frac{1}{\zeta_k}-1$. It can be shown that $\frac{\theta_{k,j}}{\phi_k}\sim Be^{\prime}(a,b)$~\cite{Armagan:GeneralBeta}; the same relationship holds for other $\mathcal{TPB}$ variables. % not clear what 'other tpb variables' means?
This relationship implies the following hierarchical structure for the loading matrix $\*\Lambda$:
\begin{eqnarray}
\gamma &\sim& \mathcal{G}(f,\nu)\label{eq:gamma-spec}\\
\eta &\sim& \mathcal{G}(e,\gamma)\\
\tau_k &\sim& \mathcal{G}(d,\eta)\\
\phi_k &\sim& \mathcal{G}(c,\tau_k)\label{eq:phi-spec}\\
\delta_{k,j} &\sim& \mathcal{G}(b,\phi_k)\\
\theta_{k,j} &\sim& \mathcal{G}(a,\delta_{k,j})\\
\Lambda_{k,j} &\sim& \mathcal{N}(0,\theta_{k,j}), \label{eq:lambda-spec}
\end{eqnarray} 
where the parameter $\eta$ controls the global shrinkage, $\phi_k$ controls the factor-specific shrinkage, and $\theta_{k,j}$ controls the local shrinkage for each element of the factor loading matrix $\*\Lambda$.
%\vspace{-0.3cm}

\subsection{Mixture of sparse and dense factors}\label{sec:mix-SFA}
We will define a \emph{sparse factor} as factor associated with a loading vector $\Lambda_k$ that contains one or more zeros (or minimal contribution from some number of features); we similarly define a \emph{dense factor} as a factor associated with a loading vector $\Lambda_k$ that contains no zeros (or contributions from all features). This formulation of the model (Equation~\ref{eq:lambda-spec}) makes it suitable for generating sparse factors and, simultaneously, eliminating unnecessary factors. If we removed the local sparse components, and instead let each element of the loading matrix be generated from the factor-level variance term directly, $\Lambda_{k,j} \sim \mathcal{N}(0,\phi_k)$, the model generates dense factors and simultaneously eliminates unused factors. 
%Readers may argue that such a defined dense factor does not have to be dependent though $\phi_k$ directly, for example, even for $\Lambda_{k,j}\sim P(\Lambda_{k,j}|\theta_{k,j})$, if $\phi_k$ is tuned to be big enough (recall that the shrinkage amount is inversely correlated with $\phi_k$, see table \ref{Tab:param-effect}), a dense factor can still be obtained. In our case however, such a density need $\phi_k$ to be near impossibly big to be achieved, leave aside the numerical precision limit for such a huge value, the iterations needed to converge to this number itself is formidable. 
Although there are other possible ways to model dense factors in this framework, we have found that this approach is both computationally tractable and numerically stable.

Using this approach, we added a mixture model to the prior on $\*\Lambda$ in order to jointly model both sparse and dense factors. In particular, we mix over generating each $\theta_{k,j}$ parameter from the gamma prior to encourage sparsity within a factor loading vector, and setting $\theta_{k,j}$, $j\in \{1,\dots,p\}$, to the factor-specific parameter $\phi_k$ to encourage dense factor loadings:
\begin{equation}
\theta_{k,j}\sim \pi\mathcal{G}a(a,\delta_{k,j}) + (1-\pi) \delta(\phi_k),
\end{equation}  
where $\delta(\cdot)$ is the dirac delta function, which sets $\theta_{j,k} = \phi_k$ for all $j \in \{1,\dots,p\}$.  

Let $Z\in\{0,1\}^K$ be a latent vector that indicates whether a factor is a sparse or a dense component. These indicator variables have a Bernoulli distribution with parameter $\pi$, which we further assume are generated according to a beta distribution with parameters $\alpha$ and $\beta$. Therefore, we may view the gene expression data as being generated from the following model: 

\begin{align}
\pi |\alpha,\beta &\sim Be(\alpha,\beta)\\
%\Theta_j|G_0 &\sim G_0,\ j=\{0,1\}.\\
Z_k|\pi &\sim \operatorname{Bern}(\pi),\  k=\{1,\dots, K\}\\
%\Lambda_{k,j}|Z_k &\sim p(\Lambda_{k,j}|\Theta_{Z_k})\\
\Lambda_{k,j}|Z_k &\sim \left\{ \begin{array}{ll}
p(\Lambda_{k,j}|\theta_{k,j},\delta_{k,j},\phi_k) & \mbox{if $Z_k = 1$};\\
p(\Lambda_{k,j}|\phi_{k}) & \mbox{if $Z_k = 0$}. \end{array} \right.\\
X_{i,k} &\sim \mathcal{N}(0,1)\\
Y_{i,j}|\Lambda_{k,j},X_{i,k},\psi_j &\sim \mathcal{N}\left(\sum_{k=1}^K X_{i,k} \Lambda_{k,j},\psi_j\right).
\end{align}
%instead of p(x) put the distribution Ga(x) --  

\section{Approximate inference via EM}\label{sec:EM-main}
We present a fast expectation maximization (EM) algorithm for parameter estimation in this model; we also derived a Gibbs sampler (Appendix~A).
In the Expectation step of the EM algorithm, we take expectations of the latent factors $\*X$ and latent variables $\*Z$; this is simple because $\*X$ and $\*Z$ are conditionally independent of each other with respect to $\*\Lambda$. 
We use maximum a posteriori (MAP) estimates for parameters in the M-step as in the original paper on EM~\cite{Dempster:1977} (see Appendix B for complete description). 
The posterior probability is written as follows:
\begin{align}
p(\*{\Lambda,X,Z,\Theta|Y}) &\propto p(\*{Y|\Lambda,X,\Theta,Z})p(\*{X|\Theta})p(\*{\Lambda|\Theta,Z})p(\*{Z|\Theta})p(\*\Theta)\\ 
&\propto\left[\prod_{i=1}^n \N(Y_i | \*\Lambda, X_i) \N(X_i|0,\*I_K)\right] \left[\prod_{k=1}^K\prod_{j=1}^p \N(\Lambda_{k,j} | \phi_k) ^{\mathds{1}_{Z_k=0}}\right]\notag\\
&\times \left[ \prod_{k=1}^K\prod_{j=1}^p \left\{
\N(\Lambda_{k,j} | \theta_{k,j})\Ga(\theta_{k,j} | a,\delta_{k,j})\Ga(\delta_{k,j}|b,\phi_k)
\right\}^{\mathds{1}_{Z_k=1}} \right] \notag\\
&\times \left[ \prod_{k=1}^K \mathcal{B}ern(Z_k|\pi) \right] \left[\prod_{k=1}^K \Ga(\phi_k|c,\tau_k)\Ga(\tau_k | d,\eta)\right]\notag\\
&\times \Ga(\eta | e,\gamma)\Ga(\gamma|f,\nu)\mathcal{B}eta(\pi|\alpha,\beta) \notag
\end{align}
%\begin{align}
%p(\*{\Lambda,X,Z,\Theta|Y}) &\propto p(\*{Y|\Lambda,X,\Theta,Z})p(\*{X|\Theta})p(\*{\Lambda|\Theta,Z})p(\*{Z|\Theta})p(\*\Theta)\\ 
%&\propto\left[\prod_{i=1}^n \N(Y_i | \*\Lambda, X_i) \N(X_i|0,\*I_K)\right] \left[\prod_{k=1}^K\prod_{j=1}^p \N(\Lambda_{k,j} | \phi_k) ^{\mathds{1}_{Z_k=0}}\right]\notag\\
%&\times \left[ \prod_{k=1}^K\prod_{j=1}^p \left\{
%\N(\Lambda_{k,j} | \theta_{k,j})\Ga(\theta_{k,j} | a,\delta_{k,j})\Ga(\delta_{k,j}|b,\phi_k)
%\right\}^{\mathds{1}_{Z_k=1}} \right] \notag\\
%&\times \left[ \prod_{k=1}^K \mathcal{B}ern(Z_k|\pi) \right] \left[\prod_{k=1}^K \Ga(\phi_k|c,\tau_k)\Ga(\tau_k | d,\eta)\right]\notag\\
%&\times \Ga(\eta | e,\gamma)\Ga(\gamma|f,\nu)\mathcal{B}eta(\pi|\alpha,\beta) \notag
%\end{align}
%\begin{align}\label{eq:full-like}
%p(\*{\Lambda,X,Z,\Theta|Y}) &\propto p(\*{Y|\Lambda,X,\Theta,Z})p(\*{X|\Theta})p(\*{\Lambda|\Theta,Z})p(\*{Z|\Theta})p(\*\Theta)\\ 
%&\propto \N(\mathbf{Y | \Lambda, X})\N(\*X|0,\*I) \left[\prod_{k=1}^K\prod_{j=1}^p \N(\Lambda_{k,j} | \phi_k) ^{\mathds{1}_{Z_k=0}}\right]\notag\\
%&\times \left[ \prod_{k=1}^K\prod_{j=1}^p \left\{
%\N(\Lambda_{k,j} | \theta_{k,j})\Ga(\theta_{k,j} | a,\delta_{k,j})\Ga(\delta_{k,j}|b,\phi_k)
%\right\}^{\mathds{1}_{Z_k=1}} \right] \notag\\
%&\times \left[ \prod_{k=1}^K \mathcal{B}ern(Z_k|\pi) \right] \left[\prod_{k=1}^K \Ga(\phi_k|c,\tau_k)\Ga(\tau_k | d,\eta)\right]\notag\\
%&\times \Ga(\eta | e,\gamma)\Ga(\gamma|f,\nu)\mathcal{B}eta(\pi|\alpha,\beta) \notag
%\end{align}
Key elements of EM include: 1) the posterior of $\Lambda_{k,j}$ has a normal distribution, with its mode being a function of a weighted sum of the sparse and dense components, 2) the posterior of $\theta_{k,j}$ and $\phi_k$ are in a Generalized Inverse Gaussian ($\!{GIG}$) distribution, with MAP estimates of their modes being a closed form solution to a quadratic function; however, $\theta_{k,j}$ is only associated with the sparse components, whereas $\phi_k$ is a function of both sparse and dense components. 3) The parameters $\delta_{i,k},\tau_k$ have a gamma distribution, for which the MAP estimates have a closed form solution because of conjugacy. For parameters $\phi_k, \eta$, we used their MLE estimates when MAP estimates $\equiv 0$, which is the case for the horseshoe parameterization of the $\TPB$ prior ($a = b = 0.5$). %4) The Expectation step includes the expected value of $\*X$ and the latent variable $\*Z$. %We note that since $Z \sim \operatorname{Bern}(V)$, we normalize its values to [0,1]. %move 4 to 1

\section{Stability statistics}\label{sec:robust} % change robustness and robust to stability throughout paper
Factor models suffer from unidentifiability: in the general model, the likelihood is invariant up to orthogonal rotation and scaling of the factors and loadings, and the factors and loadings may be jointly permuted without affecting the likelihood, called the \emph{label switching} problem. Because of these invariances, it is difficult to compare the results from fitted factor models, specifically $\*\Lambda$ and $\*X$. However, it is important to be able to compare these fitted matrices because we would like to, for example, quantify how well simulated data are recapitulated or evaluate how sensitive the EM algorithm is to random starting points. In the sparse matrix setting, by imposing significant sparsity on the loading matrix, we eliminate rotational invariance for the most part. We therefore construct a stability measure to compare two sparse matrices that is invariant to scale and label switching. In the dense matrix setting, we develop a stability measure that quantifies the similarity between two matrices based on their underlying basis, which is invariant to rotation, scaling, label switching, and even a varying number of recovered factors. %Note that identifiability problem for the dense factor is trivial compare to the sparse factors, as for the later,we need $\*\Lambda$ itself to be identifiable since it represent the classification of the samples, while for the former, the confounding are captured in $\*\Lambda \*X$, which is invariant to orthogonal rotations.  % I don't know what the point of this last sentence is?   

\subsection{Stability statistic for sparse factors}
\label{sec:robust-sparse}
We propose the following stability measurement for two sparse matrices. Let $K_1$, $K_2$ be the number of rows for two sparse matrix $\*\Lambda_1$ and $\*\Lambda_2$, let $\Sigma \in [0,1]^{K_1\times K_2}$ denote the correlation matrix generated from two fitted sparse matrices $\hat{\*\Lambda}_1 \in \Re^{K_1 \times p}$ and $\hat{\*\Lambda}_2 \in \Re^{K_2 \times p}$ by computing the absolute value of the pairwise Pearson's correlations among each sparse matrix column, we consider the following statistic: % fix this notation, rows/columns -- the dimensions of the matrices need to be spelled out. Same for the other section on dense matrices too.
\begin{align}
r_s&= \frac{1}{2K_1}\sum_{l=1}^{K_1}\left\{\max(|\Sigma_{l,.}|)-\frac{\sum_{t=1}^{K_2}I(|\Sigma_{l,t}| > |\overline{\Sigma_{l,.}}|)\Sigma_{l,t}}{K_2-1}\right\}\\ 
&+ \frac{1}{2K_2}\sum_{t=1}^{K_2} \left\{\max(|\Sigma_{.,t}|)-\frac{\sum_{l=1}^{K_1}I(|\Sigma_{l,t}|>\overline{|\Sigma_{.,t}|})\Sigma_{l,t}}{K_1-1}\right\}\notag
\end{align}
where $\overline{|\Sigma_{l,.}|}$ and $\overline{|\Sigma_{.,j}|}$ denote the mean for the $i^{th}$ row and the $j^{th}$ column.
The idea behind this metric is as follows: given two sparse matrices that are perfect matches despite label switching,
there should be exactly one $\Sigma_{i,j}=1$ for the $i$th row and $j$th column, and the rest should be closer to zero (although, because we do not enforce orthogonal factor loadings, probably not exactly zero). The stability measure $r_s$ should reward this scenario, but penalize the comparisons when there are zero or more than one $\Sigma_{i,j}\approx 1$ for the $i^{th}$ row or $j^{th}$ column (\emph{factor splitting}). Conversely, we do not want to penalize small correlations among factors as correlations may exist, so we only penalize correlations that are greater than the mean correlation value for that factor, which may be smaller than the correlation between matching factors (with correlation near one) and larger than the correlation between non-matching factors (with correlation closer to zero). 

\subsection{Stability for dense factors}\label{sec:robust-dense}
We built a stability measure to quantify the similarity of two dense matrices based on the estimates of the covariance of the features of matrix $\*M$, $\*M^T \*M$, Although $\*M$ itself is unidentifiable up to an orthogonal rotation, the form $\*M^T \*M$ is identifiable, so we will compare two dense matrices with these features using their respective covariance matrices. The problem of comparing two covariance matrices $\*\Sigma_1 = \*M_1^T \*M_1$ and $\*\Sigma_2 = \*M_2^T \*M_2$ has been well studied~\cite{Anderson:2003}. A test statistic that is a function of the determinant of the two covariance matrices will quantify the difference between the two~\cite{Anderson:2003,Li:2012} for example.
A determinant-based approach was rejected, though, because, in our model $p \gg K$, so the $p \times p$ covariance matrices are singular and therefore will have a determinant of zero. To address this, a simple squared trace, $Tr(\*\Sigma_1-\*\Sigma_2)^2$, was recently proposed to measure the distance between two dense matrices~\cite{Li:2012}.
%Based on the independent and identically distributed samples drawn from the two covariance matrices, $\*X_1 \sim \mathcal{N}(\mu_1,\*\Sigma_1)$ and $\*X_2 \sim N(\mu_2,\*\Sigma_2)$, they further proposed a test statistic $t \sim \mathcal{N}(0,1)$. We note that our problem is a simple version in the sense that instead of the observed samples drawn from the distribution, our dense matrix $\*M$ is the design matrix that directly forms the covariance matrix, that is, for a dense matrix $\*M \in \Re^{k \times p}$, our $\*\Sigma \in \Re^{p \times p} = \*M^T \*M$ is known. % I didn't get most of this. perhaps we should write it out a bit more slowly?
This metric is rotation invariant, invariant to label switching, and allows singular matrices; to make it scale invariant, we scale each row of the original matrices by $\left(M_{i,.}-\frac{1}{p}\sum_{j=1}^p M_{i,j}\right)\left(\frac{1}{p-1}M_i^T M_i\right)^{-1/2}$. Given two scaled dense matrices $\*M_1 \in \Re^{K_1 \times p}$ and $\*M_2 \in \Re^{K_2 \times p}$, we compare them by using the trace squared: % take out lambda, put in a general dense matrix.
\begin{equation}
r_d=\frac{1}{p^2}Tr(\*M_1^T\*M_1-\*M_2^T\*M_2)^2,
\end{equation}
which is proportional to the distance between the two matrices, with smaller values representing greater similarity in this scenario.
%We also apply this metric to the estimated dense factor matrices $\*X$. % what are the characteristics of this metric? Invariant up to rotation, scaling, label switching, etc? What is r_d\in ?? add all of these things in here.

\section{Results}\label{sec:result}
\subsection{Simulated data}\label{sec:sim-result}
\begin{figure}[!htb] 
\begin{center}   
\includegraphics[width=1\linewidth]{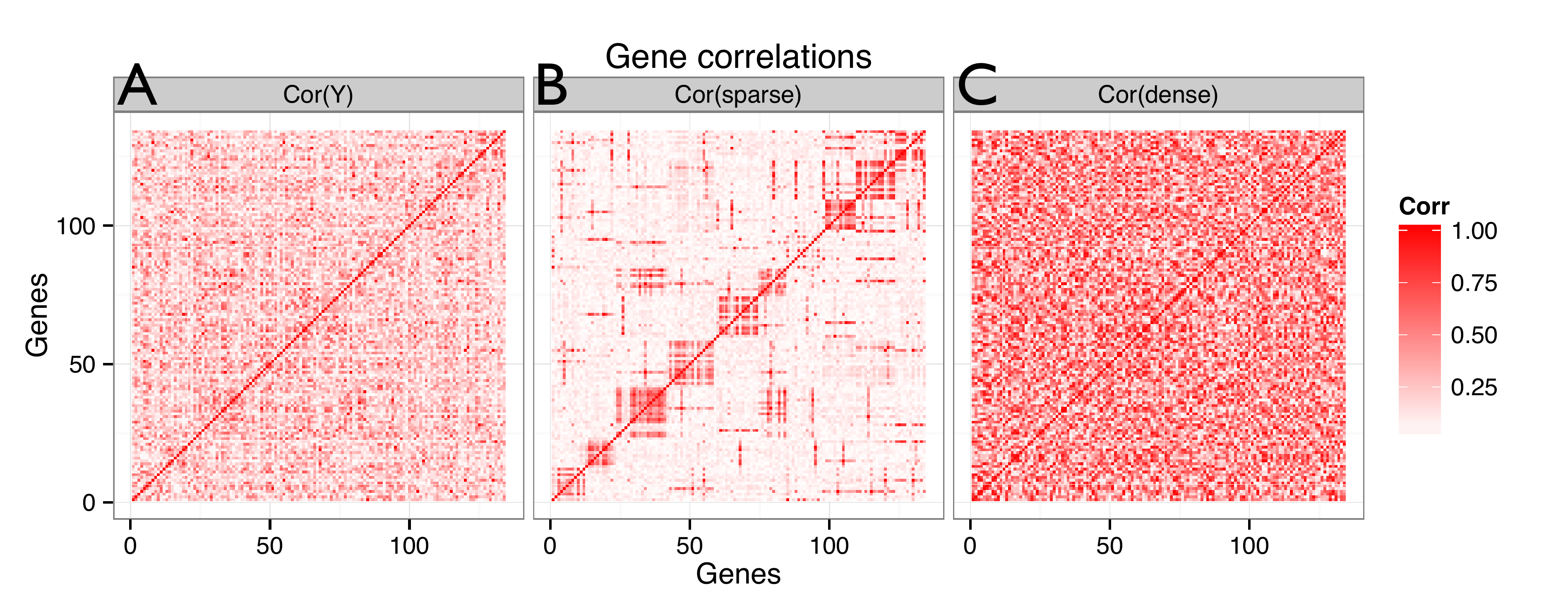}
\caption{{{\bf Correlation patterns within the simulated data.} The absolute value of the Pearson correlation coefficient between all pairs of genes is shown as a heatmap, where the rows and columns of each matrix correspond to genes. Panel A: $\*Y^T\*Y$, Panel B: $\*\Lambda^T\*\Lambda$, and Panel C: $\Omega^T\Omega$. }}\label{fig:simulation}
\end{center}
\end{figure}

To test the performance of this model, we simulated two types of gene expression measurements. First, we simulated ten data sets with only sparse components, in consideration of the models in this comparison that do not handle confounders explicitly (Sim 1). Second, we simulated ten data sets with sparse components plus dense confounders (Sim 2). The two types of simulated data were generated from the following model:
\begin{equation}
\*Y=\*X\*\Lambda + \*F\*\Omega + \*\epsilon
\end{equation}
where $\*\Lambda$ and $\*\Omega$ correspond to the sparse and dense loading matrices, and $\*X$ and $\*F$ correspond to the sparse and dense factors, respectively. To generate $\*\Lambda \in \Re^{n \times p}$, for each row of $\*\Lambda$, We sampled the number of genes in a single sparse factor from $Unif[10,20]$ and then assigned values from $\N(0,1)$ at random to the included genes and set the loadings for the excluded genes to zero. The indices of the included genes were randomly sampled from the $p$ total genes. Some genes may appear in multiple rows, and thus correlations among the factors are possible. We sampled each row of $\*X_i\sim \mathcal{N}(0,\*I_K)$. 
% Under this generative model, the sparse genes $\*{Y_s} = \*X \*\Lambda \sim \mathcal{N}(0,\*\Lambda^T \*\Lambda)$. % this is only true in expectation, not in general!
We simulated the dense matrix $\Omega_{i,j} \sim \mathcal{N}(0,1)$ $i=1,\dots,n$, $j=1,\dots,p$, and the error term $\epsilon_j \sim \mathcal{N}(0,1)$. 
For Sim 1, we simulated ten sparse factors; for Sim 2, we added five dense factors. For both simulations, we used these simulated factors to generate a gene expression matrix with dimension $n=200$, $p=500$ (Figure~\ref{fig:simulation}). The same simulation scheme was replicated ten times to generate ten data sets for both Sim 1 and Sim 2.

\subsection{Related methods for comparison}

We validated our model on these simulations, and compared the results against six other related methods. We ran KSVD~\cite{Elad:2006KSVD}, BFRM~\cite{Carvalho:BFRM}, SPCA~\cite{Witten:2009}, SBIF~\cite{Bhat:biometrika}, our model, after controlling for PCs in the $\*Y$ matrix (SFAmix2), and our model, leaving $\*Y$ as it was given (SFAmix), in the following way.

We ran KSVD assuming one element in each linear combination, which best recapitulated the sparsity in the simulations for both Sim 1 and Sim 2. The default number of iterations was used. We gave the method the correct number of factors. We also ran the method on Sim 1 setting $K=20$. We ran BFRM and SPCA with the correct factor numbers; we used default values for the other parameters. We also ran BFRM and SPCA on Sim 1  setting $K=20$. For SBIF, maximum a posteriori (MAP) estimates of $\*X$ and $\*\Lambda$ were used as the final point estimates. This method selected the factor number nonparametrically; However, we seeded the method with the correct number of factors. For Sim 1, we also seeded SBIF with $K=20$.

For SFAmix2, we ran SFAmix as below, but controlled for confounders in matrix $\*Y$ before applying our model to the residuals from a fitted linear model with the original matrix $\*Y$ and the first five principal components (PCs) of $\*Y^T\*Y$. % doubel check this. 
For SFAmix, we initialized the program with $50$ factors, and we set the parameter values to $a=b=c=d=0.5$ and $\nu=1$ to recapitulate the horseshoe. We set $\alpha=\beta=1$ for a uniform prior distribution on the mixture proportions. We assessed convergence by checking changes in the number of non-zero elements $l = \sum_{k=1}^K ||\*\Lambda_k||_0$ in each iteration, and stopped when $l$ was stable for $20$ iterations. We also ran SFAmix using the correct number of factors on Sim 1.

Since a number of methods in this comparison did not recover matrices with substantial sparsity, we post processed the results for these methods to determine the sparse and dense loadings. We chose a cutoff $t$, different for each method, so that, for a factor loading $k$, we thresholded the vector elements to count the number of non-zero features in that factor: $l_k=\sum_{j=1}^p \mathds{1}\{|\Lambda_{k,j}|>t\}$. We determined this cutoff based on factor loading histograms, resolving ambiguous cutoff levels in favor of the correct number of sparse and dense factors. Then we set elements for which $|\Lambda_{k,j}|\le t$ to zero. For SFAmix, we used the posterior probability of the $Z_k$ variables to determine whether a factor was sparse or dense (with a naive cutoff of $0.5$). We found for SFAmix that the threshold for removing a feature from a factor $t$ was $<10^{-10}$, requiring minimal post-processing to determine the gene clusters.

\subsection{Comparison between six methods on simulated data}
\begin{figure}[!htb] 
\begin{center}   
\includegraphics[width=0.7\linewidth]{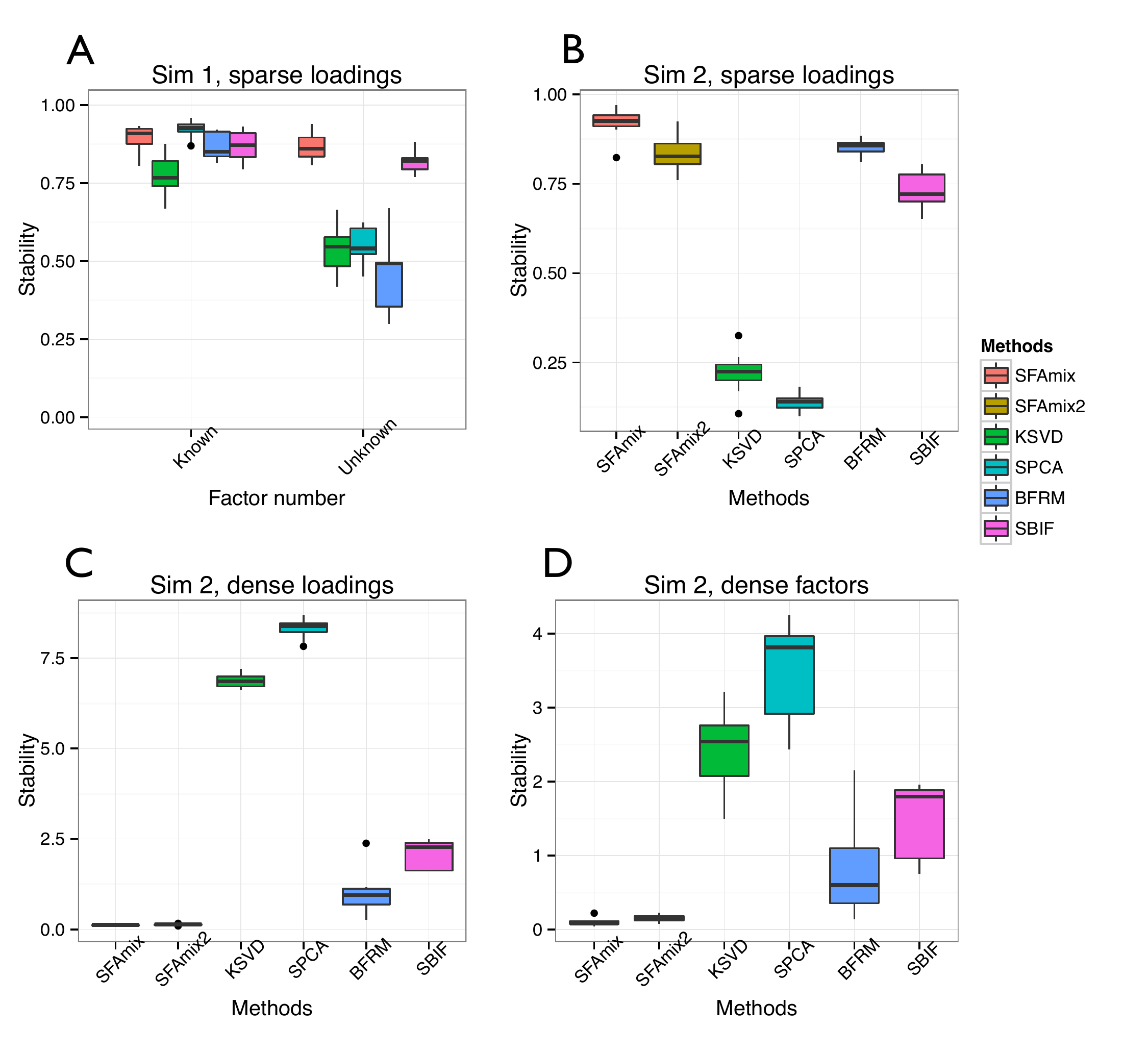}
\caption{{{\bf Stability measures for sparse and dense matrices over ten simulations.} Panel A: Average $r_s$ for sparse loadings on Sim 1, for both known and unknown numbers of factors. Panel B: Average $r_s$ for sparse loadings on Sim 2. Panel C: Average $r_d$ for dense loadings on Sim 2. Panel D: Average $r_d$ for dense factors on Sim 2.}}\label{fig:robust-all-sparse}
\end{center}
\end{figure}

% all results should be in past tense. give our method an acronym!
We compared our mixture factor analysis model, SFAmix, and our model with a two-stage approach, SFAmix2, to KSVD, BFRM, SPCA, and SBIF. We evaluated the performance of each method based on the stability statistics between the true simulated and the recovered latent spaces, for both sparse and dense loadings and factor matrices.  
We ran each of the five methods on the ten data sets in Sim 1, and we compared each recovered sparse factor loadings $\hat{\Lambda}$ with the true loading matrix $\*\Lambda$ (Figure~\ref{fig:robust-all-sparse}). When the correct factor number was known for all methods other than SFAmix, all methods were able to recover the sparse factor loadings well, all producing an average stability measure $r_s > 0.75$ over the ten simulations. When the factor number was unknown (SFAmix was given $K=50$ and all other methods were given $K=20$, for simulated $K=15$) SFAmix recovered the sparse loading matrix equally well, followed by SBIF, while the remaining methods performed substantially worse. This suggests a benefit of the nonparametric behavior of SFAmix and SBIF, which both estimated the number of factors effectively when the underlying factor number was unknown \emph{a priori}.

For Sim 2, we found that SFAmix recovered the sparse loadings well, followed by BFRM, SFAmix2, SBIF, KSVD, and SPCA (Figure~\ref{fig:robust-all-sparse}B). Indeed, SFAmix was able to recover both the sparse and dense loading matrices without knowing the number or proportion of sparse and dense factors beforehand (Figure~\ref{fig:robust-all-sparse}C). The amount of post processing required for BFRM may have artificially inflated the quality of those results relative to SBIF in particular. BFRM and SBIF allow variability on the shrinkage applied across factors; thus, they recover matrices with confounding factors better than KSVD and SPCA, which impose equal shrinkage across factors (Figure~\ref{fig:robust-all-sparse}B). This difference is reflected in the dense stability measure, where SFAmix and SFAmix2 had the smallest average distance between the recovered and the true covariance matrices, followed by BFRM, SBIF, KSVD and SPCA (Figure~\ref{fig:robust-all-sparse}C). We used the dense stability metric to compare the recovered factors corresponding to the dense loadings to the original dense factors, and we find an identical ranking of methods in terms of the factor recovery but with substantially greater variance across the different data sets in Sim 2 (Figure~\ref{fig:robust-all-sparse}D). The results for Sim 2 suggest that estimating the sparse and dense components jointly offers benefits over the two-stage method (SFAmix2), which, even given the correct factor numbers, performs worse than the joint model in recovering the sparse components (Figure~\ref{fig:robust-all-sparse}B,D). 

% we never ran the r_d for the factors? any difference in results from loadings? not much difference according to r_d on dense factors.
% Huh? K=15? That is never listed in the section on how you ran the methods! Shouldn't it be initialized at the simulated number of latent factors, not the *average* simulated number of factors? ** all simulations have 15 factors.

\begin{figure}[!htb] 
\begin{center}   
\includegraphics[width=1\linewidth]{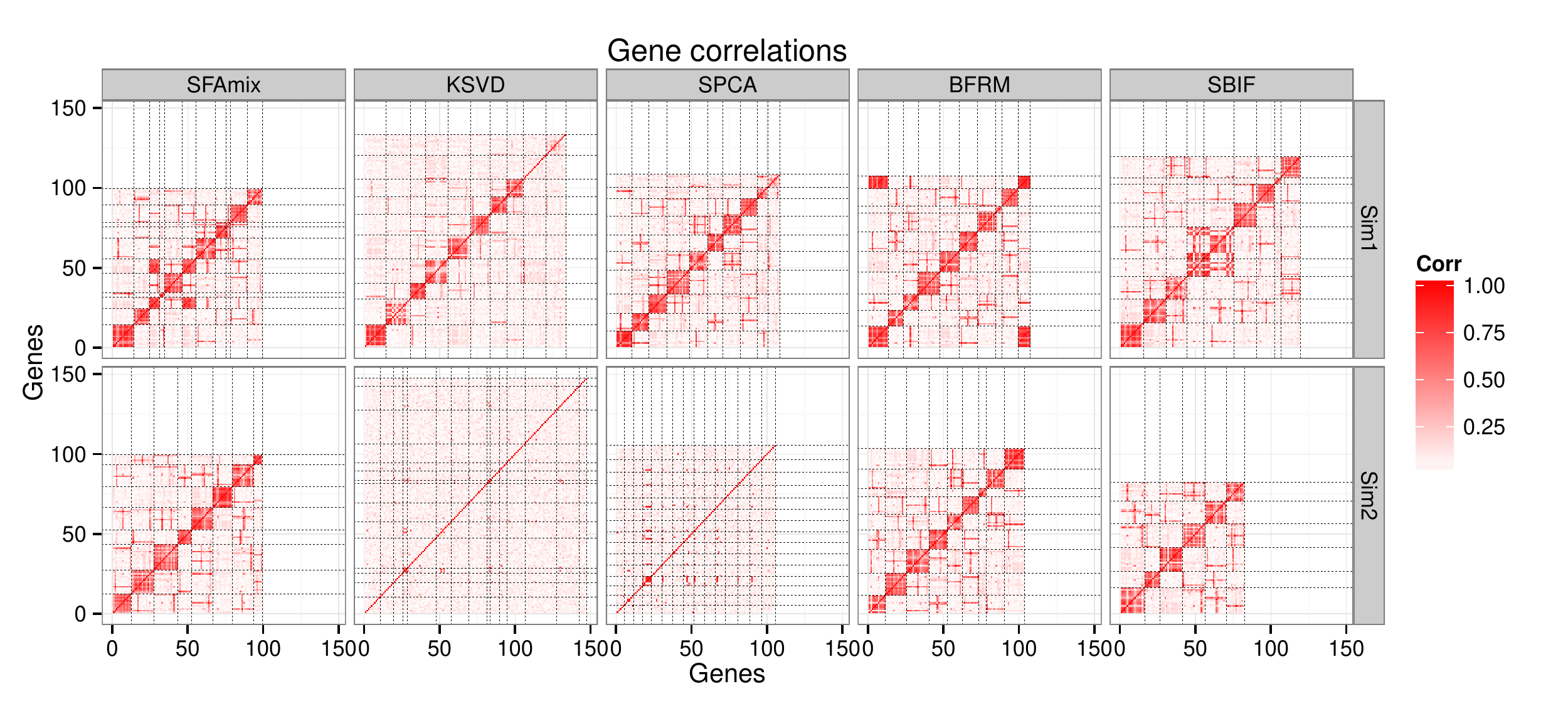}
\caption{{{\bf Absolute value of the correlation between genes based on the recovered factor loadings across five methods.} The top and bottom panels show $\*\Lambda^T\*\Lambda$ for the recovered sparse loading matrices $\*\Lambda$ for Sims 1 and 2, respectively, across the five methods; the x- and y-axes are the genes included in those factors. Correlations between genes in a single factor are partitioned by black lines.}}\label{fig:gene-cor}
\end{center}
\end{figure}

We further investigated the recovered gene clusters in the sparse loadings for both Sim 1 and 2. We found that, for Sim 1, SFAmix and SPCA recapitulated the level of sparsity in the simulated loadings; in particular, the average number of non-zero components in a sparse loading ($l_k$) for SFAmix, KSVD, SPCA, BFRM and SPIF were 10, 50, 23, 495, and 500 respectively, where the simulated average cluster size was 15 (Figure~\ref{fig:gene-cor}). 
For Sim 2, we found that the sparsity levels were recovered well by SFAmix, and also by BFRM and SBIF when the number of sparse and dense loadings were approximated correctly (Figure~\ref{fig:gene-cor}). KSVD and SPCA do not approximate the sparse clusters well in the presence of dense factors. SFAmix recover the sparse latent structure well relative to other methods in the presence of confounding factors, with minimal post-processing.
% thinking through this a bit: Since *you* chose the cutoff t, it's hard to criticize a method for not producing equivalent sparsity when this level is arbitrary?  ** how about this? we let the computer choose the cutoff automatically, for our method, it reached a cutoff that computer can't handle any more, so it's automatically set to zero. other methods can't reach such a threshold. It's not fair for our method to say that we choose a cutoff, because we actually don't, the values are set to zero. Or, we can say our method shrink values to exactly 0. for other methods, we have to choose a cutoff for them that's above the computer precision.  

%\vspace{-0.2cm}

% a few sentences on the computational time need to be added.

\subsection{Gene expression study}\label{sec:real}
An RNA microarray generates gene transcription levels for tens of thousands of genes from an RNA sample rapidly and at low cost. Biologists have shown that genes are not transcribed into mRNA as independent units, but instead as correlated components of biological networks with various biochemical roles~\cite{Kim:2011,Chen:2008}.
As a result, genes that share similar biological roles may have correlated expression levels across samples because, for example, they may be regulated for a similar cellular purpose by a common transcription factor that is expressed at different levels across samples. Identifying these correlated sets of genes from high dimensional gene expression measurements is a fundamental biological problem~\cite{Allocco:2004,Dupont:2012,Marbach:2012} with many downstream applications.

\subsubsection{Latent factors recovered from the gene expression data}\label{sec:real-result}
We applied our method to expression levels from $8,718$ genes measured in a sample of $480$ human immortalized blood cell lines (LCLs)~\cite{Mangravite:2013}. The data were processed according to previous work~\cite{Mangravite:2013}; however, neither known covariates nor PCs were controlled for before quantile normalization. We also removed genes with probes on the gene expression array that aligned to multiple regions of the genome using a BLAST analysis and human reference genome hg19. In this experiment, the number of correlated sets of genes may be large relative to the number of genes in the gene expression matrix (and, certainly, relative to the number of samples) if we indeed identify small clusters of co-regulated genes. We set $K=4000$ and ran EM from ten starting points with $a=b=c=d=0.5$, $\nu=1$ and $\alpha=\beta=1$. We recovered approximately $350$ factors across different random starting values, approximately 25-30 of which were dense factors. %need to be clear about results across random runs. ** r_s for different runs \in [0.3,0.33], because of factor missing and splitting. 

\begin{figure}[!htb] 
\begin{center}   
\includegraphics[width=1\linewidth]{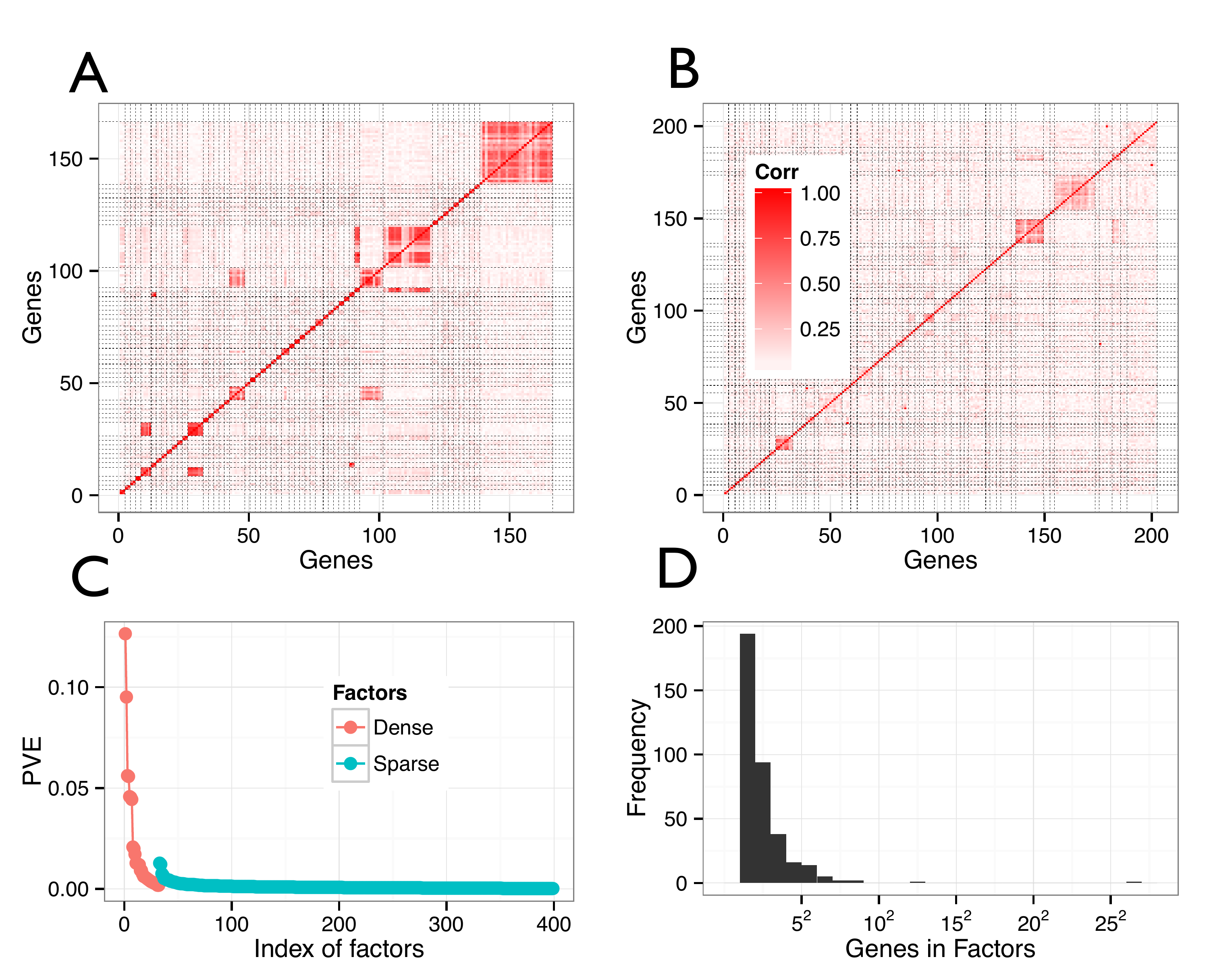}
\caption{{{\bf Gene correlation patterns, proportion of variance explained, and the number of genes in each sparse factor.} Panel A: Absolute value of Pearson's correlation between genes associated with sparse factors as a heatmap for factors with average correlation $>0.4$ and cluster size $<50$. Genes in a single factor are partitioned by black lines. Panel B: Same heatmap for 50 factors selected at random with average correlations $<0.4$ and cluster size $<50$. Panel C: Percentage of variance explained by both dense and sparse factors (ordered). Panel D: Histogram of the number of genes in each sparse factor.}}\label{fig:lam-real}
\end{center}
\end{figure}

We present results from the run that produced the most factors. For this run, we found a total of $399$ factors, of which $32$ were dense (Figure~\ref{fig:lam-real}A,B). We found that 98\% of the sparse factors contained fewer than $50$ genes, and 81\% contained fewer than $10$ genes (Figure~\ref{fig:lam-real}D).
To quantify gene correlation patterns within each factor, we calculated the correlation matrix using the gene expression values in the residual matrix $\mathbf{Y-\Omega F}$ for each gene included in each sparse factor (Figure~\ref{fig:lam-real}A,B). We found that our model recovered factors containing groups of strongly correlated genes, even when the correlation was confounded by the structure of the dense factors in the original matrix $\*Y$.

\begin{figure}[!htb] 
\begin{center}   
\includegraphics[width=0.6\linewidth]{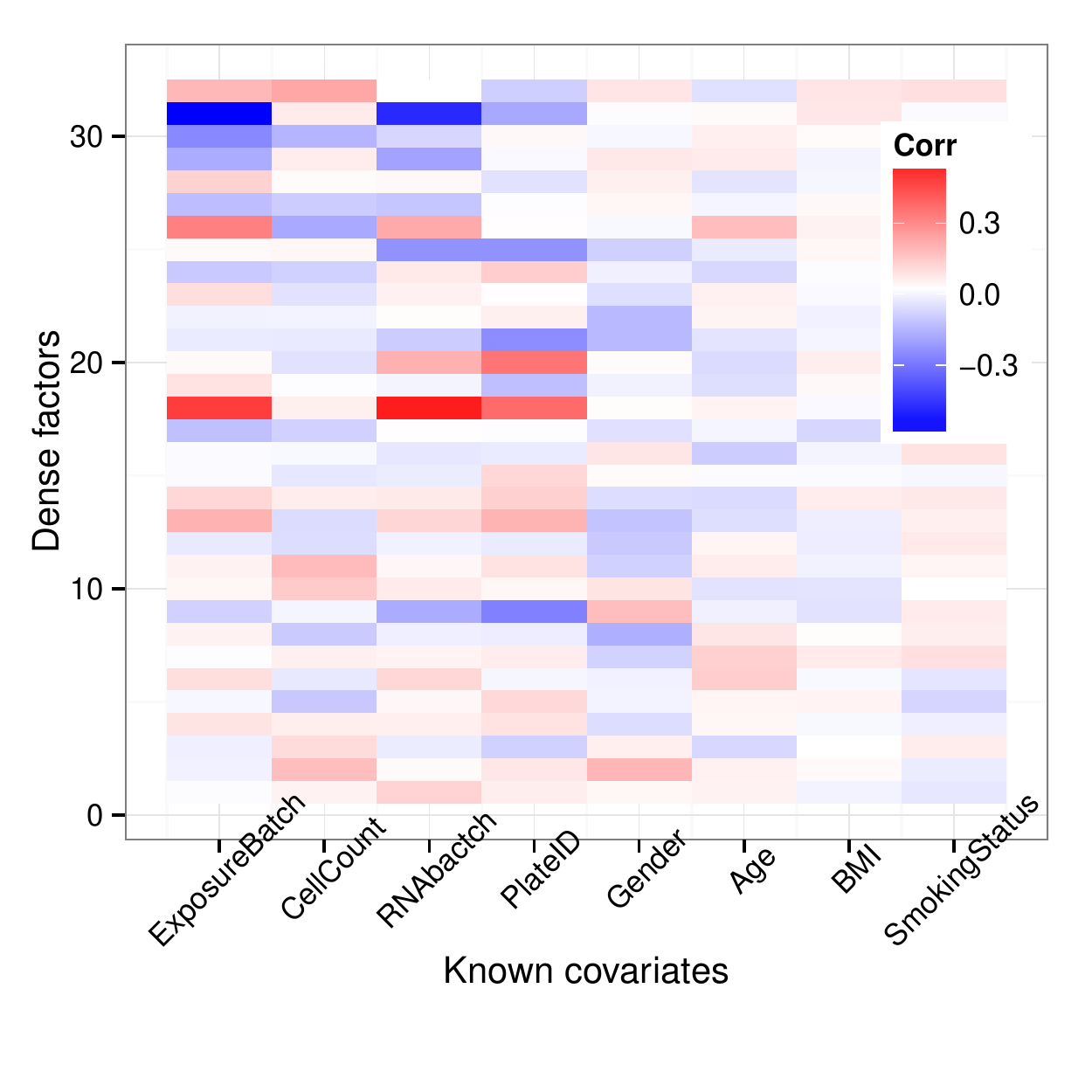}
\caption{{{\bf Correlations between the recovered dense factors and known covariates.} The known covariates are plotted along the $x$-axis and the recovered dense factors are shown on the $y$-axis; colors represent various levels of Pearson's correlation coefficient between each covariate and the recovered factors.}}\label{fig:cor_dense}
\end{center}
\end{figure}

A further look at the proportion of variance explained (PVE; Figure~\ref{fig:lam-real}C) shows that dense factors individually explain as much as 13\% of the total variance in the gene expression matrix. The sparse factors individually explain as much as $\approx$ 1.3\% of the total variance, which is more than some of the dense factors explain. One feature of our joint mixture model is that sparse factors may capture substantial PVE, instead of controlling for this variance through PCs in a two-stage approach (SFAmix2) or implicitly controlling sparsity via a decreasing prior on the PVE (SBIF). Furthermore, we found that the recovered dense factors correlated well with some known biological and technical covariates, including batch effects, which are known to cause a substantial amount of variation in gene expression levels~\cite{Leek:2010} (Figure~\ref{fig:cor_dense}).% what do you want to say about this? can you also plot a figure of average correlation vs PVE or # genes (or both, in two different figures) so we can see if thre is a correlation there? ** ave corr and # genes has relationship, ave corr and PVE hasn't. see plot.

%\begin{figure}[!htb] 
%\begin{center}   
%\includegraphics[width=1\linewidth]{PV_Ngene_aveCor.pdf}
%\caption{{{\bf Gene correlation pattern, proportion of variance explained and distribution of the number of genes in each factor.}.}}\label{fig:lam-real}
%\end{center}
%\end{figure}

% only showing the best ones is a little rough. Can you instead show the best in one matrix and the worst in the another (as ranked by average correlation)?

\subsubsection{Ontology term enrichment validates recovered gene clusters}\label{sec:eqtl}
Genes that have correlated transcription levels often have similar molecular functions~\cite{Kim:2011,Chen:2008}. To validate the gene clusters recovered by SFAmix, we applied the Gene Ontology enrichment analysis tool, DAVID~\cite{huang:2009}, to the genes in each sparse factor. Using FDR $\leq 0.2$, we found that $145$ sparse factors were enriched for $1,917$ different biological functions (Supplemental Table~S1). For example, a sparse factor with $39$ genes, including CMPK2, DDX60, and SP110, was enriched for the GO terms \emph{response to virus} (FDR $\leq 3.11\times 10^{-13}$) and \emph{antiviral defense} (FDR $\leq 3.26\times 10^{-8}$). The substantial enrichment of GO terms in the recovered sparse factors suggests that the induced gene clusters recovered by this model are biologically meaningful. Furthermore, these specific GO terms indicate that these samples have mounted a coordinated cellular response to virus, which, as we discuss later, reflects the immortalization process for the specific type of cells in this study~\cite{Caliskan2011}. %minal & yoav's paper.

\subsubsection{eQTL analysis finds pleiotropic eQTLs}\label{sec:eqtl}

One downstream application of identifying subsets of correlated genes is to find genetic variants that are associated with transcription levels of the recovered subsets of possibly co-regulated genes~\cite{Subramanian:2005,Zhernakava:2013,Suissa:2009,Xiong:2012}. 
% cite, and let's list a few more downstream applications of IDing gene networks -- somatic mutations involved in cancer, rare variants, enriched differential gene expression data, etc. list and cite, briefly.
To further validate the recovered gene clusters, we performed eQTL association mapping to the sparse factors to identify genetic variants that regulate the corresponding small gene clusters (\emph{pleiotropic eQTLs}). For this experiment, we projected each recovered factor to the quantiles of a standard normal distribution across the samples; we then tested for associations between each of these normalized latent factors and $\approx$ 2.6 million genetic variants (genotyped in the same individuals) using univariate Bayesian tests~\cite{Servin:2007}. We also ran the same association tests on the permuted normalized latent factors to compute the false discovery rate (FDR) for specific $\log_{10}BF$ values.
We identified $7,154$ associated genetic variants (FDR$\leq 0.2$; $\log_{10}BF \geq 3.70$), and $5,568$ associated genetic variants at a more strict FDR (FDR $\leq 0.05$; $\log_{10}BF \geq 4.37$); all identified eQTLs are presented in Supplemental Table~S2.

\begin{figure}[!htb] 
\begin{center}   
\includegraphics[width=0.8\linewidth]{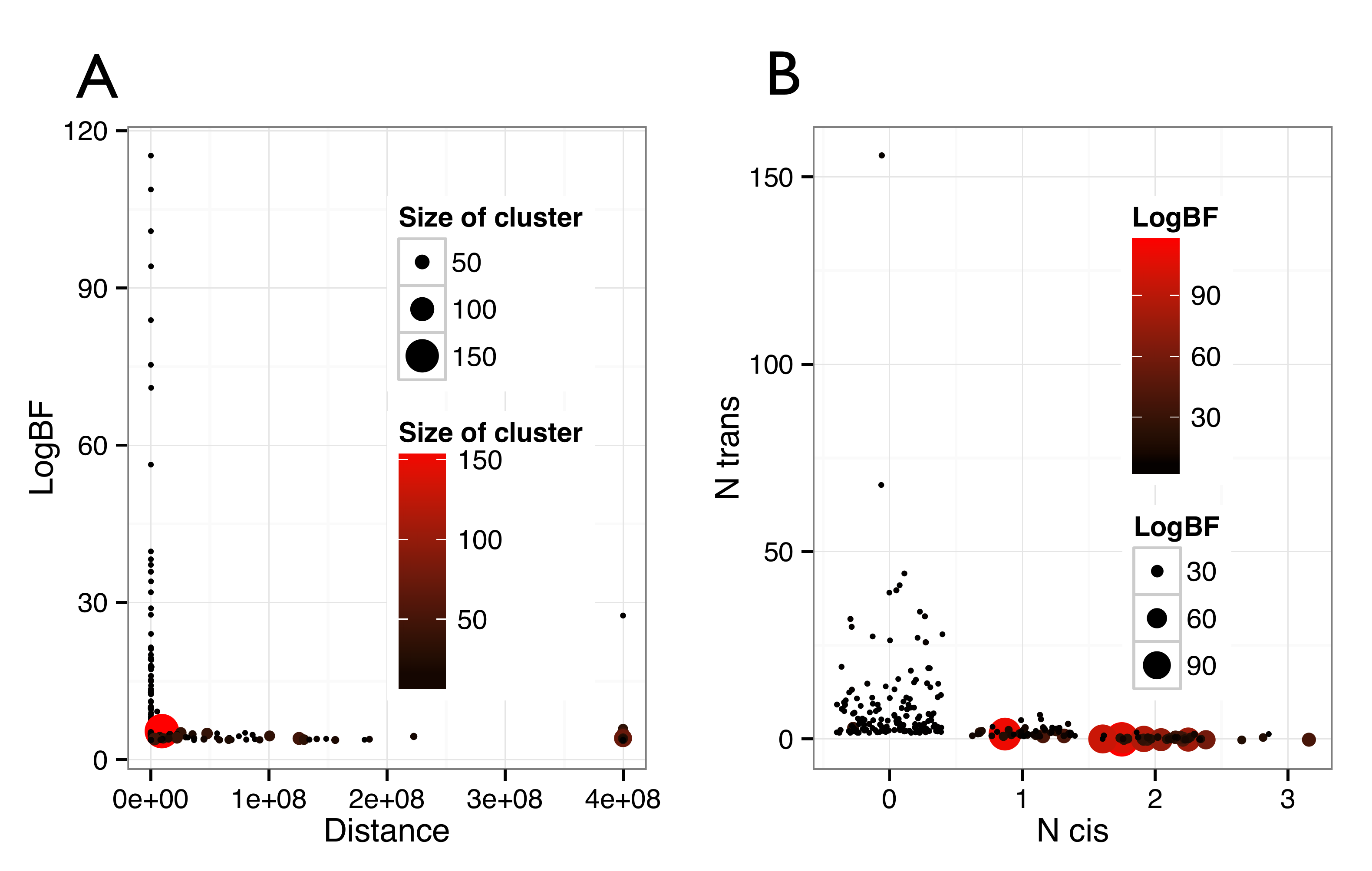}
\caption{{{\bf Distance of the associations, cluster size and $\log_{10}BF$ for the top associations above FDR of 0.2.}} Panel A: x-axis shows distance between the SNPs and their associated factors (the smallest distance to the TSS or TES of any gene within that factor); the y-axis corresponds to the $\log_{10}BF$ association values; the size of the points corresponds to the size of the gene clusters. The distance $4\times 10^8$ represents SNPs located on different chromosomes from all of the genes in the associated cluster. Panel B: For all factors, the number of $cis$ gene-SNP associations is shown on the x-axis and the the number of $trans$ gene-SNP associations is shown on the y-axis; The size of points correspond to the $\log_{10}BF$ values.}\label{fig:cis-trans} 
\end{center}
\end{figure}

We found that $257$ out of $367$ of our sparse factors ($70\%$) had at least one eQTL (FDR$\leq 0.2$). We define \emph{cis} associations as variants located within 1 Mb of the transcription start site (TSS) or the transcription end site (TES) of any gene in the factor. We found $5,318$ ($76\%$) cis-associations of $7154$ total associations, recapitulating previous studies showing many more significant cis-eQTLs than trans-eQTLs~\cite{Stranger:2005,Mangravite:2013}. If we consider only the most significantly associated eQTL for each factor, $95$ out of $257$ factors with eQTLs (37\%) are in \emph{cis}; however, this proportion goes up to 60\% ($86$ out of $143$) at an FDR of $0.05$ ($\log_{10}BF \geq 4.37$), and 84\% ($72$ out of $86$) at an FDR of $0.01$ ($\log_{10}BF \geq 5.17$), suggesting that the \emph{cis} associations represent stronger genetic effects than the \emph{trans} associations~\cite{Grundberg2012}. All associations with $\log_{10}BF \geq 30$ have a cluster size of less than or equal to three genes; generally the eQTL is a short distance from the cis-gene's transcribed region (Figure~\ref{fig:cis-trans}A). Less significant associations ($\log_{10}BF \leq 30$) show more variability in distance to the closest gene and cluster size (Figure~\ref{fig:cis-trans}A). As the number of genes in \emph{cis} to the eQTL for a cluster increased, the association significance also tended to increase (Figure~\ref{fig:cis-trans}B). These associations suggest that this type of factor model can be used to capture small groups of genes that appear to be co-regulated by pleiotropic genetic loci.

% need to compute FDR here and say how many signicant eQTLs there are at FDR \leq 0.2
For comparison, we found $119$ genetic variants associations with the dense factors (FDR$\leq 0.20$), and $22$ associated genetic variants at a stricter FDR (FDR$\leq 0.05$). This proportion of dense factors with eQTLs is smaller than the genetic associations for the sparse factors, supporting the hypothesis that most of the dense factors are not genetically driven but represent biological and experimental confounders. This also suggests that a joint modeling of gene clusters and confounding effects does not remove genetic signal unintentionally, although it is possible that a genetic effect constitutes only a small proportion of the variance explained by a dense factor, so those factors would still not appear to be associated with genetic variants.

Association mapping identified eQTLs associated with two factors, the first including ten genes (DDX58, GMPR, IFIT2, IFIT3, IFIT5, MOV10, OASL, PARP12, PARP9, XAF1), the second including four genes (CD55, CR1, CR2, IFNA2).  Both eQTLs are unlinked with (i.e., in trans) all of the genes included in the factors. Both of these factors are enriched for GO terms related to interferon response, or response to invasion of host cells by pathogens including viruses and tumor cells; the first factor is enriched for \emph{interferon-induced 56K protein}  (FDR $\leq 3.45\times 10^{-04}$), and the second factor is enriched for \emph{Sushi4 domain} (FDR $\leq 5.42\times 10^{-3}$) that is activated in response to specific viruses including Epstein-Barr. Both of these factors are relevant to the cell type in this study, lymphoblastoid cell lines, which have been immortalized using the Epstein-Barr virus, and it appears that we are able to observe the response that these cells have mounted against the viral pathogen. For the first factor, the trans-eQTL is located within a K-lysine acetyltransferase (KAT8, also known as MYST1), which is in our gene expression data but not included in this factor, and is a known interferon effector gene~\cite{Fusco:2013}. The eQTL for the second factor is similarly located within the TRAPPC9 gene, not in our gene expression data set, which is in the NF-$\kappa$B pathway and is activated during viral stress of host cells.
%\begin{figure}[!htb] 
%\begin{center}   
%\includegraphics[width=1\linewidth]{Figure5_1.pdf}
%\includegraphics[width=1\linewidth]{Figure5_2.pdf}
%\includegraphics[width=1\linewidth]{Figure5_3.pdf}
%\includegraphics[width=1\linewidth]{Figure5_4.pdf}
%\caption{{{\bf The $\log_{10}BF$ from the eQTL analysis on one factor running across a total of approximately 2.3 million SNP variants.} The Quantile-Quantile plot for the $\log_{10}BF$ and the empirically obtained $\log_{10}BF$ are plotted on the right.}}\label{fig:eqtl-1} 
%\end{center}
%\end{figure}

\begin{figure}[!htb] 
\begin{center}   
\includegraphics[width=0.6\linewidth]{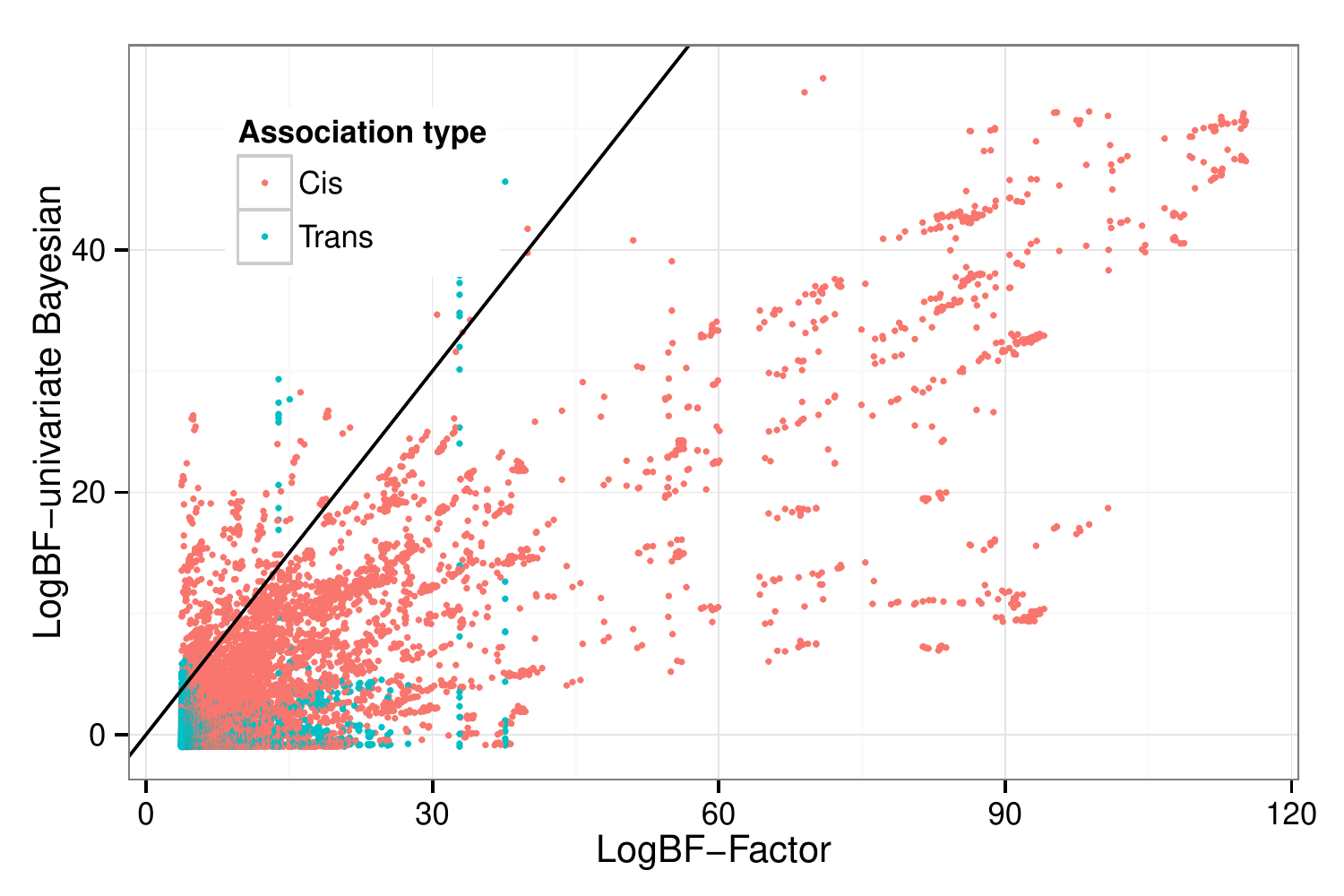}
\caption{{{\bf The $\log_{10}BF$ for SNP-factor associations and univariate SNP-gene associations for the genes within each factor.}} The x-axis shows the $\log_{10}BF$ for the SNP-factor associations, and the y-axis corresponds to the $\log_{10}BF$ of SNP-gene associations within factors. The black line represents perfect correlation. Only SNP-factor associations with an FDR $\geq 0.2$ were used.}\label{fig:uni-factor} 
\end{center}
\end{figure}

We also performed a univariate Bayesian test for association between the genes within each factor and the SNPs associated with these factors for SNP-factor associations with an FDR$\leq 0.20$. We found that by jointly testing for association with the clustered genes, we identified associations with greater significance than testing the genes separately (Figure~\ref{fig:uni-factor}).  
%\vspace{-0.5cm}

\section{Conclusions}
%\vspace{-0.2cm}
We developed a model for sparse factor analysis using a three parameter beta prior to induce shrinkage in the loading matrix at three levels of the hierarchy: globally, factor-specific, and element-wise. We found that this model has favorable properties for estimating possibly high-dimensional latent spaces, including accurate recovery of sparse signals and a non-parametric property of removing unused factors. We extended this model to explicitly include dense factors by adding a two-component mixture model within this hierarchy. We developed two simple metrics for stability across sparse and dense matrices that are invariant to scale, label switching, and (for dense matrices) orthogonal rotation. We validated our model on simulated data, and showed that our model recapitulated both sparse and dense factors with high accuracy relative to current state-of-the-art methods. We applied our model to a large gene expression data set and found biologically meaningful clusters of genes. The recovered dense factors correlate well with known biological and technical covariates. We used the sparse factors to identify genetic variants that are associated with transcriptional regulation of the genes within the individual sparse factors, and our results suggest that our sparse gene clusters capture genes that are co-regulated by genetic variants, and that our method is useful for identifying pleiotrophic eQTLs.

\appendix
\section{Posterior distribution for the parameters}\label{sec:post}

The posterior probability for our model given matrix $\*Y$ is written as follows:
\begin{align}\label{eq:full-like}
p(\*{\Lambda,X,Z,\Theta|Y}) &\propto p(\*{Y|\Lambda,X,\Theta,Z})p(\*{X|\Theta})p(\*{\Lambda|\Theta,Z})p(\*{Z|\Theta})p(\*\Theta)\\ 
&\propto\left[\prod_{i=1}^n \N(Y_i | \*\Lambda, X_i) \N(X_i|0,\*I_K)\right] \left[\prod_{k=1}^K\prod_{j=1}^p \N(\Lambda_{k,j} | \phi_k) ^{\mathds{1}_{Z_k=0}}\right]\notag\\
&\times \left[ \prod_{k=1}^K\prod_{j=1}^p \left\{
\N(\Lambda_{k,j} | \theta_{k,j})\Ga(\theta_{k,j} | a,\delta_{k,j})\Ga(\delta_{k,j}|b,\phi_k)
\right\}^{\mathds{1}_{Z_k=1}} \right] \notag\\
&\times \left[ \prod_{k=1}^K \mathcal{B}ern(Z_k|\pi) \right] \left[\prod_{k=1}^K \Ga(\phi_k|c,\tau_k)\Ga(\tau_k | d,\eta)\right]\notag\\
&\times \Ga(\eta | e,\gamma)\Ga(\gamma|f,\nu)\mathcal{B}eta(\pi|\alpha,\beta). \notag
\end{align}

The conditional probability for $X_i$ has the following form:
\begin{align}
X_i|Y_i,\*\Theta &\propto 
\exp 
\left\{\sum_{i=1}^{n}
\left( 
\frac{1}{2}  (Y_i-X_i\*\Lambda)^T\*\Psi^{-1}(Y_i-X_i\*\Lambda)
\right) 
\right\} 
\exp\left\{\sum_{i=1}^{n}
\left( 
\frac{1}{2}X_i^T X_i
\right)
\right\}
\\
&\propto
\exp 
\left\{\sum_{i=1}^{n}  
\left( 
\frac{1}{2}  (X_i-\mu_{x_i})^T \*\Sigma_X^{-1}(X_i-\mu_{x_i})
\right) 
\right\}.\notag
\end{align}

Thus, we have the following conditional probability for $X_i$:

\begin{eqnarray}
X_i|Y_i \sim \mathcal{N}(\mu_{x_i}, \*\Sigma_X)\label{eq:post-x}
\end{eqnarray}
where
\begin{eqnarray}
%\mathbf{\langle X \rangle} &=& \mathbf{(\Lambda^T\Psi^{-1}\Lambda+I_K)^{-1}\Lambda^T\Psi^{-1}Y}\\
\mu_{x_i} &=& Y_i^T\*\Psi^{-1}\*\Lambda\mathbf{(\Lambda\Psi^{-1}\Lambda^T+I_K)^{-1}}\label{gib:ex}\\
\mathbf{\*\Sigma_X} &=& \mathbf{(\Lambda\Psi^{-1}\Lambda^T+I_K)^{-1}}
\end{eqnarray}
%For other parameters, the likelihood function of $\mathbf{\Lambda}$ conditional on the sparse and dense indicator variable $Z_k$ and its hierarchical parameters set $\mathbf{\Theta}$, can be written as
%\item

The conditional probability for $Z_k$ has a Bernoulli distribution:
\begin{align}
p(Z_k=1|\*\Theta) = \frac{\pi\prod_{j=1}^p\N(\Lambda_{k,j} | \theta_{k,j})\Ga(\theta_{k,j} | a,\delta_{k,j})\Ga(\delta_{k,j}|b,\phi_k)}{(1-\pi)(\prod_{j=1}^p \N(\Lambda_{k,j} | \phi_k))+\pi\prod_{j=1}^p\N(\Lambda_{k,j} | \theta_{k,j})\Ga(\theta_{k,j} | a,\delta_{k,j})\Ga(\delta_{k,j}|b,\phi_k)}
\end{align}

Let $\rho_k = p(Z_k=1|\*\Theta)$; then the conditional probability for $Z_k$ is 

\begin{align}
Z_k|\*\Theta &\sim \!{B}ern(\rho_k)
\end{align}

To derive the conditional probabilities for the parameters generating the matrix $\*\Lambda$, we note that many of them have a generalized inverse Gaussian distribution, conditional on $Z_k$:

If $Z_k = 1$
\begin{align}
\Lambda_{k,j}|\*Y,\*X,\Theta_{k,j},\psi_{j,j} &\sim \N\left(
\frac{\frac{1}{\psi_{j,j}}\sum_{i=1}^n(y_{i,j}-\sum_{\tilde{k}\neq k}
  x_{i,\tilde{k}}\Lambda_{\tilde{k},j})x_{i,k}}{\frac{1}{\psi_{j,j}}\sum_{i=1}^n
x_{i,k}^2 + \frac{1}{\theta_{k,j}}},\frac{1}{\psi_{j,j}}\sum_{i=1}^n
x_{i,k}^2 + \frac{1}{\theta_{k,j}}
\right)\\
\theta_{k,j}|\Lambda_{k,j},\delta_{k,j} &\sim \mathcal{GIG}\left(a-\frac{1}{2},2\delta_{k,j},\Lambda_{k,j}^2\right)\\
\delta_{k,j}|\theta_{k,j},\phi_{k} &\sim \mathcal{G}(a+b,\theta_{k,j}+\phi_k)\\
\phi_k|\delta_{k,j},\tau_{k} &\sim \mathcal{G}\(pb+c,\sum_{j=1}^p\delta_{k,j}+\tau_k\)
\end{align}

If $Z_k = 0$
\begin{align}
\Lambda_{k,j}|\*Y,\*X,\phi_k,\psi_{j,j} &\sim \N\left( 
\frac{\frac{1}{\psi_{j,j}}\sum_{i=1}^n\(y_{i,j}-\sum_{\tilde{k}\neq k}
  x_{i,\tilde{k}}\Lambda_{\tilde{k},j}\)x_{i,k}}{\frac{1}{\psi_{j,j}}\sum_{i=1}^n
x_{i,k}^2 + \frac{1}{\phi_k}},\frac{1}{\psi_{j,j}}\sum_{i=1}^n x_{i,k}^2 + \frac{1}{\phi_k}
\right)\\
\phi_k|\tau_{k},\Lambda_{k,j} &\sim  \mathcal{GIG}\left(c-\frac{p}{2},2\tau_k,\sum_{j=1}^p \Lambda_{k,j}^2\right),
\end{align}

where we used $\tilde{k}\neq k$ to denote any element but element $k$.

The following parameters are not sparse or dense component specific, and they each have a gamma conditional probability because of conjugacy:
\begin{align}
\tau_k|\phi_k,\eta &\sim \mathcal{G}\(c+d,\phi_k+\eta\)\\
\eta|\gamma,\tau_k &\sim \mathcal{G}\(Kd+e,\gamma+\sum_{k=1}^K\tau_k\)\\
\gamma|\eta,\nu &\sim \mathcal{G}(e+f,\eta+\nu).
\end{align}

The mixing proportion $\pi$ has a beta conditional probability: 
\begin{align}
\pi|\alpha,\beta,Z_k \sim \mathcal{B}eta\(\alpha+\sum_{k=1}^K \mathds{1}_{Z_k=0},K-\sum_{k=1}^K\mathds{1}_{Z_k=1}+\beta\),
\end{align}
where $\mathds{1}$ is the indicator function.

Finally, the variance of the error term has an inverse gamma distribution: 
\begin{align}
\psi_{j,j}|\*Y,\*X,\*\Lambda \sim \mathcal{IG}\left(\frac{n}{2}-1,\frac{\sum_{i=1}^n \left(y_{i,j}-\sum_{k=1}^K x_{i,k}\Lambda_{k,j}\right)^2}{2} \right).
\end{align}

%\end{enumerate}

\section{Expectation maximization algorithm}\label{sec:EM}
We describe an expectation maximization algorithm, where we take the expected values of the latent variables $Z$ and $\*X$, enabling conjugate gradient methods for point estimates of the parameters in this space. To derive the EM updates, we write the auxiliary function, using the expected complete log posterior probability in lieu of the likelihood, $Q(\*\Theta)=\^{\ell_c(\*\Theta, \*\Lambda|\*Z,\*X,\*Y)}$ as:
\begin{align}
Q(\*\Theta) &\propto 
\sum_{i=1}^n\sum_{j=1}^p\^{\log p(y_{i,j}|\*\Lambda,\*X,\*\Theta,\*Z)}+\sum_{i=1}^n\sum_{k=1}^K \^{\log p(x_{i,k}|\*\Theta)}\\
&+\sum_{k=1}^K\sum_{j=1}^p\^{p(Z_k|\*\Theta)\log p(\Lambda_{k,j}|\*\Theta,Z_k)}+\log p(\*\Theta)\notag\\
&\propto
-\frac{p}{2}\ln|\*\Psi|-\sum_{i=1}^n\sum_{j=1}^p\frac{\(y_{i,j}-\sum_{k=1}^K\^{x_{i,k}}\Lambda_{k,j}\)^2}{2\psi_{j,j}}-\sum_{i=1}^n\sum_{k=1}^K\frac{\^{x_{i,k}^2}}{2}\notag\\
&+\sum_{k=1}^K\sum_{j=1}^p
\^{1-\^{z_k}}\left\{
-\frac{1}{2}\ln\phi_k-\frac{\Lambda_{k,j}^2}{2\phi_k}
\right\}\notag\\
&+\sum_{k=1}^K\sum_{j=1}^p
\^{z_k}\left\{
-\frac{1}{2}\ln\theta_{k,j}-\frac{\Lambda_{k,j}^2}{2\theta_{k,j}}
+a\ln\delta_{k_j}+(a-1)\ln\theta_{k_j}-\delta_{k,j}\theta_{k,j}
\right\}\notag\\
&+\sum_{k=1}^K\sum_{j=1}^p\^{z_k}\left\{b\ln\phi_k+(b-1)\ln\delta_{k,j}-\phi_k\delta_{k,j}\right\}\notag\\
&+\sum_{k=1}^K\left\{\^{z_k}\ln\pi+(1-\^{z_k})\ln(1-\pi)\right\}\notag\\
&+\sum_{k=1}^K\left\{
c\ln\tau_k+(c-1)\ln\phi_k-\tau_k\phi_k+d\ln\eta+(d-1)\ln\tau_k-\eta\tau_k
\right\}\notag\\
&+e\ln\gamma+(e-1)\ln\eta-\gamma\eta+f\ln\nu+(f-1)\ln\gamma-\nu\gamma +\alpha \ln \pi + \beta \ln (1-\pi)\notag
\end{align}

%\begin{enumerate}
%\item
First we write out the equations for the three expected sufficient statistics identified above. In section \ref{sec:post}, we established that $X_i|Y_i$ has a Gaussian distribution; then the expected value $\^{X_i|Y_i}$ is computed in the E-step as: 
\begin{align}
\^{X_i|Y_i} &= Y_i^\*T\*\Psi^{-1}\*\Lambda\mathbf{(\Lambda\Psi^{-1}\Lambda^T+I_K)^{-1}}
\end{align}
%\item

Similarly, $\^{x_{i,k}^2}$ is computed as follows:
\begin{align}
\^{x_{i,k}^2}&=\Sigma_{X_{k,k}}+\^{x_{i,k}}^2\\
\^{x_{i,\tilde{k}}x_{i,k}}&=\^{x_{i,\tilde{k}}}\^{x_{i,k}}+\Sigma_{X_{\tilde{k},k}}.
\end{align}

% YOU also need E{\*X^TX | \*Y}!!!, *** Yes, added in below before the MAP for \lambda_{k,j}

The expected value of $Z_{k}|\*\Theta$ was derived in Section \ref{sec:post} as:
\begin{align}
\langle z_{k}|\*\Theta\rangle&= p(Z_k=1|\*\Theta)\\
&=\frac{\pi\prod_{j=1}^p\N(\Lambda_{k,j} | \theta_{k,j})\Ga(\theta_{k,j} | a,\delta_{k,j})\Ga(\delta_{k,j}|b,\phi_k)}{(1-\pi)(\prod_{j=1}^p \N(\Lambda_{k,j} | \phi_k))+\pi\prod_{j=1}^p\N(\Lambda_{k,j} | \theta_{k,j})\Ga(\theta_{k,j} | a,\delta_{k,j})\Ga(\delta_{k,j}|b,\phi_k)}.\notag
\end{align}

The parameter updates are computed in the M-step. We obtain their MAP estimates $\hat{\*\Theta}=\argmax_\*\Theta Q(\*\Theta)$. Specifically, we solve equation $\frac{\partial Q(\*\Theta)}{\partial\*\Theta} = 0$ to find the closed form of their MAP estimates. The same updates are obtained by finding the mode of the conditional probability of each parameter, as in Appendix~\ref{sec:post}. 

Our MAP estimate for $\Lambda_{k,j}$ is a function of the weighted sum of the two variance terms $\theta_{k,j}$ and $\phi_k$: 

\begin{align}
\hat{\Lambda}_{k,j} &=
\frac{\frac{1}{\psi_{j,j}}\sum_{i=1}^n\left(y_{i,j}\^{x_{i,k}}-\sum_{\tilde{k}\neq k}
  \^{x_{i,\tilde{k}},x_{i,k}}\Lambda_{\tilde{k},j}\right)}{\frac{1}{\psi_{j,j}}\sum_{i=1}^n
  \^{x_{i,k}^2} + \frac{\z_k}{\theta_{k,j}}+\frac{1-\z_k}{\phi_k}},
\end{align}

where $\^{x_{i,k}^2}$ is calculated in the E-step. $\Sigma_{X_{k,k}}$ was derived in Appendix~\ref{sec:post} as the $(k,k)$th element in the $\*\Sigma_X$ matrix, and $\Sigma_{X_{\tilde{k},k}}$ as the $\tilde{k},k$th element in the $\*\Sigma_X$ matrix.

As shown in Appendix~\ref{sec:post}, $\theta_{k,j}$ has a generalized inverse Gaussian conditional probability, and its MAP estimates can either be obtained by directly taking the mode of this distribution, or by solving a quadratic formula. We obtain the following form for the parameter updates:
\begin{align}
\hat{\theta}_{k,j}&=\frac{2a-3+\sqrt{(2a-3)^2+8\Lambda_{k,j}^2\delta_{k,j}}}{4\delta_{k,j}}.
\end{align}

The estimates for $\hat{\delta}_{k,j}$ are trivially obtained as:
\begin{align}
\hat{\delta}_{k,j}&=\frac{a+b-1}{\theta_{k,j}+\phi_k}.
\end{align}

The estimates for $\phi_k$ are also based on a generalized inverse Gaussian. Unlike $\theta_{k,j}$, $\phi_k$ generates both sparse and dense factors, so the estimates are a function of a weighted sum of parameters from both components:
\begin{align}
\hat{\phi}_k&=\frac{h+\sqrt{h^2+\chi \omega}}{\chi}\label{eq:phi-EM},
\end{align}
where
\begin{align}
h &= pb\z_k+c-1-\frac{p}{2}(1-\z_k)\\
\chi &= 2\left(\z_k\sum_{j=1}^p\delta_{k,j}+\tau_k\right)\\
\omega &= \sum_{j=1}^p\Lambda_{k,j}^2.
\end{align}

The following parameters have similar updates to $\delta_{k,j}$, which have natural forms because of conjugacy:
\begin{align}
\hat{\tau}_k&=\frac{c+d-1}{\phi_k+\eta}\\
\hat{\eta}&=\frac{Kd+e-1}{\gamma+\sum_{k=1}^K\alpha_k}\\
\hat{\gamma}&=\frac{e+f-1}{\eta+\nu}.
\end{align}

The prior on the indicator variable for sparse and dense components, $\pi$, has a beta distribution, and its geometric mean is the following:
\begin{align}
\langle \ln \pi\rangle&=\psi\left(\sum_{k=1}^K{\z_k}+\alpha\right)-\psi\left(K+\alpha+\beta\right) \label{eq:V1-EM}
\end{align}
where $\psi$ is the digamma function.

The variance for the error term has the following update:
%\item
\begin{align}
\hat{\psi}_{j,j} = \frac{\sum_{i=1}^n \left(y_{i,j}-\sum_{k=1}^K \^{x_{i,k}}\Lambda_{k,j}\right)^2}{n}.
\end{align}
%\end{enumerate}

\section*{Acknowledgements}
The authors would like to thank Sayan Mukherjee and David Dunson for helpful conversations. All data are publicly available: the gene expression data were acquired through GEO GSE36868, and the genotype data were acquired through dbGaP, acquisition number phs000481, and generated from the Krauss Lab at the Children's Hospital Oakland Research Institute. This work was supported in part by U19 HL069757: Pharmacogenomics and Risk of Cardiovascular Disease.  We acknowledge the PARC investigators and research team, supported by NHLBI, for collection of data from the Cholesterol and Pharmacogenetics clinical trial.

%\begin{supplement}
%\sname{Supplement A}\label{suppA}
%\stitle{Supplement\_Gao\_Brown\_Engelhardt.pdf}
%\slink[url]{http://www.e-publications.org/ims/support/dowload/imsart-ims.zip}
%\sdescription{Additional details regarding our results on the CAP gene expression eQTL data.}
%\end{supplement}

%\bibliographystyle{unsrt}
\bibliographystyle{unsrt}

\bibliography{ref}
\end{document}